%% file: vwap_opt_exec.tex
\pdfoutput=1

\documentclass[12pt]{article}
\usepackage{fullpage,graphicx,psfrag,amsmath,amsfonts,verbatim}
\usepackage[small,bf]{caption}
\usepackage{algorithmicx,algpseudocode}
\usepackage{url,algorithm,amsthm}
\input defs.tex

\usepackage{subcaption}
\usepackage{lscape} 
\usepackage{booktabs}

\newcommand{\figref}[1]{\figurename~\ref{#1}}

\usepackage{hyperref} 

\usepackage{tikz}
\usetikzlibrary{decorations.pathreplacing}

\usepackage[toc,page]{appendix}
\graphicspath{{./Graphics/}}

\title{Volume Weighted Average Price Optimal Execution}
\author{Enzo Busseti \and Stephen Boyd}

\date{\today}

\begin{document}
\maketitle

\begin{abstract}
We study the problem of optimal execution of a trading order under Volume Weighted Average Price (VWAP) benchmark, 
from the point of view of a risk-averse broker. The problem consists in minimizing mean-variance of the slippage, with 
quadratic transaction costs. We devise multiple ways to solve it, in particular we study how to incorporate the 
information coming from the market during the schedule. Most related works in the literature eschew 
the issue of imperfect knowledge of the total market volume. We instead incorporate it in our model. 
We validate our method with extensive simulation of order execution on real NYSE market data. 
Our proposed solution, using a simple model for market volumes,
 reduces by $10\%$ the VWAP deviation RMSE of the standard ``static'' solution (and can simultaneously reduce
transaction costs).
\end{abstract} 


\section{Introduction}
\label{sec:introduction}
Most literature on optimal execution focuses on the Implementation Shortfall (IS) objective, 
minimizing the execution price with respect to the market price at the moment the order is submitted. 
The seminal papers \cite{bertsimas1998optimal}, \cite{almgren2001optimal} and \cite{obizhaeva2006optimal} derive the optimal schedule for various risk preferences and market impact models. 
However most volume on the stock markets is traded with Volume Weighted Average Price (VWAP) orders, benchmarked to the average market price during the execution horizon \cite{madhavan2002vwap}.
Using this benchmark makes the problem much more compelling from a stochastic control standpoint 
and prompts the development of a richer model for the market dynamics. 
The problem of optimal trade scheduling for VWAP execution has been studied originally \cite{konishi2002optimal} in a static optimization setting (the schedule is fixed at the start of the day). 
This is intuitively suboptimal, since it ignores the new information coming as the schedule progresses. 
Some recent papers \cite{humphery2011optimal} \cite{mcculloch2012mean} \cite{frei2013optimal} extend the model and incorporate the new information coming to the market but rely on the crucial assumption that the total market volume is known beforehand. 
Other works \cite{bialkowski2008improving} take a different route and focus on the empirical modeling of the market volumes.
A recent paper \cite{gueant2013vwap} studies the stochastic control problem including a market impact term, while the work by Li \cite{li2013dynamic} takes a different approach and studies the optimal placement of market and limit orders for a VWAP objective. 
Our approach matches in complexity the most recent works in the literature (\cite{frei2013optimal}, \cite{gueant2013vwap})
with a key addition: we don't assume that the total market volume is known and instead treat it as a random variable. 
We also provide extensive empirical results to validate our work.

We define the problem and all relevant variables in \S\ref{sec:probl_formulation}. 
In \S\ref{sec:static_solution} we derive a ``static'' optimal trading solution. In
\S\ref{sec:dynamic_sol} we develop a ``dynamic'' solution which uses the information coming
from the market during the schedule in the best possible way: 
as our estimate of the total market volume improves we optimize our trading activity accordingly.
In \S\ref{sec:empirical_results} we detail the simulations of trading we performed, 
on real NYSE market data, using our VWAP solution algorithms. 
We conclude in \S\ref{sec:conclusions}. 

\section{Problem formulation}
\label{sec:probl_formulation}
We consider, from the point of view of a broker, 
the problem of executing a trading order issued by
a client. 
The client decides to trade $C \in \integers_+$ shares 
of stock $k$ over the course of a market day.
By assuming $C > 0$ we restrict our analysis to ``buy''
orders. If we were instead interested in ``sell'' orders we 
would only need to change the appropriate signs.
We don't explore the reasons for the client's order (it could be
for rebalancing her portfolio, making new investments, \emph{etc.}).
The broker accepts the order and performs all the trades in the market 
to fulfill it. The broker has freedom
in implementing the order (can decide when to buy and in what amount)
but is constrained to cumulatively trade the amount $C$ over the course of the day.
When the order is submitted client and broker agree on an 
execution \emph{benchmark} price which regulates the compensation
of the broker and the sharing of risk. The broker is payed by 
the client an amount equal to the number of shares traded times the execution benchmark,
 plus fees (which we neglect).  
In turn, the broker pays 
for his trading activity in the market. 
Some choices of benchmark prices are:
\begin{itemize}
\item stock price at the start of the trading schedule. This gives rise to
\emph{implementation shortfall} execution (\cite{bertsimas1998optimal}, \cite{almgren2001optimal}),
in which the client takes no risk (since the benchmark price is fixed);
\item stock price at day close. This type of execution can misalign the broker and client
objectives. The broker may try to profit from his executions by pushing the closing price
up or down, using the market impact of his trades;
\item Volume Weighted Average Price (VWAP), the average stock price throughout the day weighted
by market volumes. This is the most common benchmark price. It encourages the broker
to spread the execution evenly across the market day, minimizing market impact and detectability
of the order. It assigns most risk associated with market price movements to the client,
so that the broker can focus exclusively on optimizing execution. 
\end{itemize}
In this paper we derive algorithms for optimal execution under the VWAP
benchmark. 

\subsection{Definitions}
\label{subsec:definitions}
We work for simplicity in discrete time.
We consider a market day for a given stock,
split in $T$ intervals of the same length. 
In the following $T$ is fixed to 390, so each interval is one minute long. 

\paragraph{Volume}
We use the word \emph{volume} to denote an integer number of 
traded shares (either by the market as a whole or by a single agent).  
We define $m_t \in \reals_{+}$ for $t= 1, \ldots, T$,
the number of shares of the stock traded by the whole market 
in interval $t$, which is non-negative. 
We note that in reality the market volumes $m_t$ are integer,
not real numbers. This approximation
is acceptable since the typical number of shares traded is much greater than 1
(if the interval length is 1 minute or more) so the integer rounding 
error is negligible. 
These market volumes are distributed according to a joint
probability distribution 
\[
  f_{m_{1:T}}(m_1, \ldots, m_T).
\]
In \S\ref{subsec:market_volume_model} we propose a model
for this joint distribution.
We also define the total daily volume
\[
V = \sum_{t=1}^T m_t
\]
We call $u_t \in \reals_+$  the number of shares
of the stock 
that our broker trades in interval $t$,
for $t= 1, \ldots, T$. (Again we assume that the 
volumes are large enough so the rounding error is negligible.)
By regulations these must be non-negative, so that
all trades performed by the broker as part of the order have the same sign.

\paragraph{Price}
Let $p_t \in \reals_{++}$ for $t= 1, \ldots, T$ be the average market price 
for the stock in interval $t$. This is defined as the VWAP of all trades over interval $t$.
(If during interval $t$ there are $N_t > 0$ trades in the market, each one with 
volume $\omega_i \in \integers_{++}$ and price $\pi_i \in \reals_{++}$, then
$
p_t = {\sum_{i=1}^{N_t} \omega_i \pi_i}/{\sum_{i=1}^{N_t} \omega_i}.
$)
If there are no trades during interval $t$ then $p_t$ is undefined and in practice
we set it equal to the last available period price.
We model this price process as a geometric random walk with zero drift. 
The initial price $p_0$ is a known constant.
Then the price increments 
$\eta_{t} \equiv \frac{p_{t} - p_{t-1}}{p_{t-1}}$
for $t = 1, \ldots, T$
are independent and distributed as 
\[
\eta_{t} \sim \mathcal{N}(0, \sigma_t),
\]
where $\mathcal{N}$ is the Gaussian distribution.
The period volatilities $\sigma_t \in\reals_+$ for $t= 1, \ldots, T$ are constants known from
 the start of the market day. 
We define the market VWAP price as
\BEQ
\label{def:vwap_price}
p_{\text{VWAP}} = \frac{\sum_{t=1}^T m_t p_t}{V}.
\EEQ

\paragraph{Transaction costs}
We model the transaction costs by introducing the  
\emph{effective} price $\hat p_t$, 
defined so that
the whole cost of the trade at interval $t$ is
$u_t \hat p_t$. Our model captures
instantaneous transaction costs, in particular
the cost of the bid-ask spread, not the cost of long-term market impact.
(For a detailed literature review on transaction costs and
market impact see \cite{bouchaud2009markets}.)
Let $s_t \in \reals_{++}$ be the average fractional (as ratio of the stock price)
bid-ask spread in period $t$.
We assume the broker trades the volume $u_t$ 
using an optimized trading algorithm that mixes optimally
 market and limit orders.
The cost or proceeding per share of a buy market order is on average
$p_t (1 + {s_t}/{2})$ while for a limit order
it is on average $p^{}_t(1 - {s^{}_t}/{2})$.
Let $u_\text{LO}$ and $u_\text{MO}$ be the portions of $u^{}_t$ 
executed via limit orders and market orders, respectively, so that
$u_\text{LO} + u_\text{MO} = u^{}_t$. We require that the algorithm uses
trades of the same sign, so $u_\text{LO}$, $u_\text{MO}$, and $u^{}_t$
are all non-negative (consistently with the constraint we introduce in
\S\ref{subsec:constraints}).
We assume that the fraction of market orders over the traded volume
is proportional to the \emph{participation rate}, defined as ${u^{}_t}/{m^{}_t}$. 
So
 \[
\frac{u_\text{MO}}{u^{}_t} = \frac{\alpha}{2}\frac{u^{}_t}{m^{}_t}
 \]
where the proportionality factor $\alpha \in \reals_+$
depends on the specifics of the trading algorithm used. 
This is a reasonable assumption, especially in the limit of small
participation rate.
The whole cost or proceedings of the trade is
\[
u^{}_t \hat p^{}_t = 
p_t \left( u_\text{LO} \left( 1 - \frac{s^{}_t}{2} \right)
+ u_\text{MO} \left( 1 + \frac{s^{}_t}{2} \right)\right)
\]
which implies 
\BEQ
\label{eq:def:trans_cost_model}
\hat p^{}_t = p^{}_t \left(  1
- \frac{s^{}_t}{2} 
+ \alpha \frac{s^{}_t}{2}\frac{u^{}_t}{m^{}_t}\right).
\EEQ
We thus have a simple model for the effective price $\hat p^{}_t$, linear in $u_t^{}$.
This gives rise to \emph{quadratic} transaction costs, a reasonable
approximation for the stock markets (\cite{bouchaud2009markets}, \cite{lillo2003master}).

\subsection{Problem objective}
\label{subsec:probl_objective}
Consider the cash flow for the broker, equal to 
the payment he receives from the client 
minus the cost of trading
\[
C p_{\text{VWAP}} - \sum_{t=1}^T u_t \hat p_t.
\]
In practice there would also be fees but we neglect them.
The trading industry usually defines the \emph{slippage} 
as the negative of this cash flow.
It represents the amount by 
which the order execution price misses the benchmark.
(The choice of sign is conventional so that the optimization problem 
consists in minimizing it). 
We instead define the slippage as 
\BEQ
 S \equiv \frac{\sum_{t=1}^T u_t \hat p_t - C p_{\text{VWAP}}}{C p_{\text{VWAP}}},
\EEQ
normalizing by the value of the order. We need this in order to compare
the slippage between different orders.
By substituting the expressions defined above we get
\begin{multline}
\label{eq:normalized_slippage}
	 S = \left.\left(\sum_{t=1}^T\left[ u_tp_t\left(1 - \frac{s_t}{2} + \alpha \frac{ s_t}{2}\frac{u_t}{m_t}\right) \right] - C\frac{\sum_{t=1}^T m_t p_t}{V}\right)\right/{C p_{\text{VWAP}}} = \\
	\sum_{t=1}^T \left[ \frac{p_t}{p_{\text{VWAP}}} \left( \frac{u_t}{C} - \frac{m_t}{V}\right)  \right] + \sum_{t=1}^T \frac{ p_ts_t}{2p_{\text{VWAP}}}\left(\alpha \frac{u^2_t}{Cm_t} - \frac{u_t}{C}\right) \simeq \\
	\sum_{t=1}^{T-1}\left[  \eta_{t+1}\left(\frac{\sum_{\tau=1}^tm_\tau}{V} - \frac{\sum_{\tau=1}^tu_\tau}{C}\right)\right] + \sum_{t=1}^T \frac{  s_t}{2}\left(\alpha \frac{u^2_t}{Cm_t} - \frac{u_t}{C}\right)
\end{multline}
where we used the two approximations (both first order, reasonable on a trading horizon of one day)
\BEA
\frac{p_{t} - p_{t-1}}{p_\text{VWAP}} &\simeq&\frac{p_{t} - p_{t-1}}{p_{t-1}} = \eta_t 
\label{eq:def:normalized_sigma}\\
\frac{p_t s_t}{p_\text{VWAP}} &\simeq& s_t. 
\label{eq:def:normalized_spread}
\EEA
We model the broker as a standard risk-averse agent, 
so that the objective function is to minimize
\[
\expect S + \lambda \var (S)
\]
for a given risk-aversion parameter $\lambda \geq 0$. These expectation
and variance operators apply to all sources of randomness in the system,
\ie, the market volumes $m$ and market prices $p$, which are independent under our model.
The expected value of the slippage is
\BEQ
	\label{eq:expected_value_slippage}
   \expect_{m, p} S = \expect_m \expect_p S = \expect_m \left[ \sum_{t=1}^T \frac{{s}_t}{2}\left(\alpha \frac{u^2_t}{Cm_t} - \frac{u_t}{C}\right)\right]
\EEQ
since the price increments have zero mean. 
Note that we leave expressed the expectation over market volumes.
The variance of the slippage is
\begin{multline}
	\label{eq:variance_slippage}
 \var_{m, p}S = \expect_{m, p} \left[\left(S - \expect_{m, p} S\right)^2\right]  = 
  \expect_{m,p} S^2 - \left(\expect_{m, p} S\right)^2 = \\
  \expect_m \expect_p S^2 - \left(\expect_m \expect_p S\right)^2 
  -\expect_m (\expect_p S)^2 + \expect_m (\expect_p S)^2 =
 \expect_{m} \var_p(S) + \var_m(\expect_{p} S).
 \end{multline}
The first term is 
\begin{multline}
\label{eq:first_term_variance}
\expect_{m} \var_p(S) = 
 \expect_m \expect_p \left[\left(\sum_{t=1}^{T-1} \eta_{t+1} 
\left(\frac{\sum_{\tau=1}^tm_t}{V} - \frac{\sum_{\tau=1}^tu_t}{C}\right)\right)^2 \right] = \\
 \expect_m \left[\sum_{t=1}^{T-1} {\sigma}^2_{t+1} \left(\frac{\sum_{\tau=1}^tm_t}{V} - \frac{\sum_{\tau=1}^tu_t}{C}\right)^2 \right]
\end{multline}
which follows from independence of the price increment.
The second term is
\BEQ
\label{eq:second_term_variance}
\var_m(\expect_{p} S) = \var_m \left( \sum_{t=1}^T \frac{ {s}_t}{2}\left(\alpha \frac{u^2_t}{Cm_t} - \frac{u_t}{C}\right)\right).
\EEQ
We drop the second term and only keep the first one, so that
the resulting optimization problem is tractable.
We motivate this by assuming, as in \cite{frei2013optimal}, that the second
term of the variance is negligible when compared to the first. This is
validated \emph{ex-post}\footnote{ 
In the rest of the paper we derive multiple ways to solve the optimization
problem of minimizing the objective \eqref{eq:objective_func_normalized}. 
For these different solution methods, the empirical value of \eqref{eq:second_term_variance} 
is between $1\%$ and $5\%$ of the value of \eqref{eq:first_term_variance}, 
 so our approximation is valid.
The results are detailed in \S\ref{subsec:aggregate_results}.}
 by our empirical studies in \S\ref{sec:empirical_results}.
We thus get
\BEQ
\label{eq:objective_func_normalized}
\expect_{m, p} S + \lambda \var_{m, p} (S) \simeq 
 \sum_{t=1}^T \expect_{m} \left[ 
 \frac{{s}_t}{2}\left(\alpha \frac{u^2_t}{Cm_t} - \frac{u_t}{C}\right) +
 \lambda {\sigma}_{t}^2 \left(\frac{\sum_{\tau=1}^{t-1}m_t}{V} - \frac{\sum_{\tau=1}^{t-1}u_t}{C}\right)^2
 \right].
\EEQ
We note that the objective function separates in a sum of terms per each time
step, a key feature we will use to apply the \emph{dynamic programming}
optimization techniques in \S\ref{sec:dynamic_sol}. 

\subsection{Constraints}
\label{subsec:constraints}
We consider the constraints that apply to the optimization 
problem. The 
optimization variables are $u_t$ for $t=  1, \ldots, T$.
We require that the executed volumes sum to the total order size C
\BEQ
\label{eq:constraint_1}
\sum_{t=1}^T u_t = C.
\EEQ
We then impose that all trades have positive sign (buys) 
\BEQ
\label{eq:constraint_2}
u_t \geq 0, \  \  t=  1, \ldots, T.
\EEQ
(If we were executing a sell order, $C<0$, we would have all $u_t \leq 0$.)
This is a regulatory requirement for institutional brokers in most markets, 
essentially as a precaution against market manipulation.
It is a standard constraint in the literature about VWAP execution.

\subsection{Optimization paradigm}
\label{subsec:optimization_paradigm}
The price increments $\eta_t$ and market volumes
$m_t$ are stochastic. The volumes
$u_t$ instead are chosen as the solution of an optimization problem.
This problem can be cast in several different ways.
We define the information set $I_t$ available at time $t$
\BEQ
\label{eq:information_set}
I_t \equiv \{(p_1, m_1, u_1), \ldots, (p_{t-1}, m_{t-1}, u_{t-1}) \}.
\EEQ
By causality, we know that when we choose the value of $u_t$ we can
use, at most, the information contained in $I_t$.
In \S\ref{sec:static_solution} we formulate the optimization problem
and provide an optimal solution for the variables $u_t$ in the case we 
\emph{do not} access anything from the information set $I_t$ when choosing $u_t$.
The $u_t$ are chosen using only information available before the trading starts. 
We call this a \emph{static} solution (or \emph{open loop} in the language of control).
In \S\ref{sec:dynamic_sol} instead we develop an optimal \emph{policy} which 
can be seen as a sequence of functions $\psi_t$ of the information set available at time $t$
\[
u_t = \psi_t(I_t).
\]
We develop it in the framework on dynamic programming and we call it \emph{dynamic} solution 
(or \emph{closed loop}).

\section{Static solution}
\label{sec:static_solution}
We consider a procedure to solve the problem described in \S\ref{sec:probl_formulation}
without accessing the information sets $I_t$. 
We call this solution \emph{static} since it is fixed at the start of the trading period. (It is computed 
using only information available before the trading starts.)
This is the same assumption of \cite{konishi2002optimal} and corresponds 
to the approach used by many practitioners.
Our model is however more flexible than \cite{konishi2002optimal},
it incorporates variable bid-ask spread
and a sophisticated transaction cost model. Still, it has an extremely 
simple numericaly solution that leverages convex optimization \cite{boyd2009convex} 
theory and software.

We start by the optimization problem with objective function 
\eqref{eq:objective_func_normalized} and the two constraints \eqref{eq:constraint_1} and
\eqref{eq:constraint_2} 
\[ 
    \begin{array}{ll}
        \text{minimize}_{u}  &  \expect_{m, p}  S + \lambda \var_{m, p} ( S)\\
        \mbox{s.t.} & \sum_{t=1}^T u_t = C\\
         & u_t \geq 0, \quad t = 1, \ldots, T.   \\ 
    \end{array}
\]
We remove a constant term from the objective and
write the problem in the equivalent form
\BEQ
    \label{eq:def_static_problem}
    \begin{array}{ll}
        \text{minimize}_{u}  
        &  
        \sum_{t=1}^T  \left[
		 \frac{ {s}_t}{2C}\left(  \alpha {u^2_t \kappa_t} - {u_t }\right) + 
 		\lambda { \sigma}_{t}^2 \left( {\left(\frac{\sum_{\tau=1}^{t-1}u_t}{C}\right)}^2 - 2M_t \frac{\sum_{\tau=1}^{t-1}u_t}{C}\right) \right]\\ 
		\mbox{s.t.}  & \sum_{t=1}^T u_t = C\\
         & u_t \geq 0, \quad t = 1, \ldots, T   \\ 
    \end{array}
\EEQ
where $M_t$ and $\kappa_t$ are the constants 
\[
M_t = \expect_{m} \left[ \frac{\sum_{\tau=1}^{t-1}m_t}{V}\right],  \quad
 \kappa_t = \expect_{m}\left[  \frac{1}{m_t}\right] 
\]
for $t = 1, \ldots, T$.
In this form, the problem is a standard quadratic program \cite{boyd2009convex} and can be solved efficiently by
open-source solvers such as ECOS \cite{domahidi2013ecos} using a symbolic convex optimization suite like 
CVX \cite{cvx} or CVXPY \cite{cvxpy}.  

\subsection{Constant spread}
We consider the special case of constant spread, $s_1 = \cdots = s_T$, which leads to a great 
simplification of the solution. 
The convex problem \eqref{eq:def_static_problem} has the form
\[
    \begin{array}{ll}
        \text{minimize}_{u} 
        &  
        \sum_{t=1}^T 
      \frac{ {s}_t}{2C}\left(  \alpha {u^2_t \kappa_t} - u_t \right)
       + \lambda \left(
       \sum_{t=1}^T { \sigma}_{t}^2 \left( U_t^2 - 2M_t {U_t}/{C}\right) \right)\equiv 
       \phi(u) + \lambda \psi(u)\\
        \mbox{s.t.}  & u\in\mathcal{C}\\
    \end{array}
\]      
where $U_t = {\sum_{\tau=1}^{t-1}u_t}/{C}$ for each $t = 1, \ldots, T$, and $\mathcal{C}$ is the convex feasible set.
We separate the problem into two subproblems considering each
of the two terms of the objective. 
The first one is
\[
\begin{array}{ll}
    \text{minimize}_u
    &  
     \phi(u)
    \\ 
    \mbox{s.t.}  & u\in\mathcal{C}\\
\end{array}
\]
which is equivalent to (since the spread is constant and $\alpha > 0$)
\[
\begin{array}{ll}
    \text{minimize}_u
    &  
    \sum_{t=1}^T 
      u^2_t \kappa_t 
    \\ 
    \mbox{s.t.}  
    & u\in\mathcal{C}\\
\end{array}
\]
The optimal solution is 
(\cite{boyd2009convex}, Lagrange duality)
\[
u_t^\star = C \frac{1/\kappa_t}{\sum_{t=1}^T 1/\kappa_t},\quad   t = 1, \ldots, T.
\]
We approximate $\kappa_t = \expect_{m}\left[  {1}/{m_t}\right] \simeq {1}/{\expect_{m}[m_t]}$ and thus
\[
u_t^\star \simeq C \frac{\expect_{m}[m_t]}{\sum_{t=1}^T \expect_{m}[m_t]} \simeq C \expect_m \left[\frac{m_t}{V}\right], \quad   t = 1, \ldots, T.  
\]
The second problem is
\[
\begin{array}{ll}
    \text{minimize}_u
    &  
    \psi(u) \equiv
    \sum_{t=1}^T { \sigma}_{t}^2 \left( U_t^2 - 2M_t {U_t}/{C}\right)\\ 
    & u\in\mathcal{C}\\
\end{array}
\]
we choose the $U_t$ such that
${ \sigma}_{t}^2 \left( U_t - M_t \right) = 0$
so $U_t = M_t$ for $t = 1, \ldots, T$.
The values of $u_1, \ldots, u_{T-1}$ are thus fixed, and we choose
the final volume $u_T$ so that $u_T = C - CU_T$. 
The first order condition of the objective function is satisfied,
and these values of $u_1, \ldots, u_T$ are feasible
(since $M_t$ is non-decreasing in $t$ and $M_T \leq1$). 
It follows that this is an optimal solution, it has values
$u_t^\star = C \expect_m \left[{m_t}/{V}\right]$ for $t = 1, \ldots, T$.

Consider now the original problem. Its objective is a convex combination (apart from a constant factor)
of the objectives of two convex problem above and all three have the same constraints set. 
Since the two subproblems share an optimal solution $u^\star$, it follows that
 $u^\star$ is also an optimal solution for the combined problem. 
Thus, an the optimal solution of (\ref{eq:def_static_problem}) in the case of constant spread
is
\BEQ
\label{eq:solutiona_static_no_spread}
u_t^\star = C \expect_m \left[\frac{m_t}{V}\right] \quad t = 1, \ldots, T.
\EEQ
This is equivalent to the solution derived in \cite{konishi2002optimal} and is the standard
in the brokerage industry. 
In our model this solution arises as the special case of constant spread,
in general we could derive more sophisticated static solutions.
We also note that we 
introduced the approximation $\kappa_t = \expect_{m}\left[  {1}/{m_t}\right] \simeq {1}/{\expect_{m}[m_t]}$.
(In practice, estimating $\expect_{m}\left[  {1}/{m_t}\right]$ would require
a more sophisticated model of market volumes than $\expect_m \left[{m_t}/{V}\right]$).
We thus expect to lose some efficiency in the optimization of the trading costs.
However, with respect to the minimization of the variance of $S$ (if $\lambda \to \infty$ or $s =0$), 
this solution is indeed optimal.
In the following we compare the performances of \eqref{eq:solutiona_static_no_spread} and of the \emph{dynamic
solution} developed in \S\ref{sec:dynamic_sol}.

\section{Dynamic solution}
\label{sec:dynamic_sol}
We develop a solution of the problem that uses all the information available at the 
time each decision is made, \ie, 
a sequence of functions $\psi_t(I_t)$ where 
$I_t$ is the information set 
available at time $t$ (as defined in \eqref{eq:information_set}).
We work in the
framework of Dynamic Programming (DP) \cite{bertsekas1995dynamic}, summarized
in \S\ref{subsec:dyn_prog}. In particular we fit our problem in the special
case of linear dynamics and quadratic costs, 
described in \S\ref{subsec:lin_quad_reg}. 
However we can't apply standard DP because the random shocks affecting
the system at different times are not conditionally independent (the market
volumes have a joint distribution). 
We instead use the approximate procedure of \cite{skaf2010shrinking}, 
summarized in \S\ref{subsec:cond_dep_dist}. 
In \S\ref{subsec:vwap_as_lqc} we finally write our 
optimization problem, defining the state, action and costs,
and in \S\ref{subsec:vwap_lqc_solution} we derive its solution. 

\subsection{Dynamic programming}
\label{subsec:dyn_prog}
We summarize here the standard formalism of dynamic programming, following \cite{bertsekas1995dynamic}. 
Suppose we have a state variable $x_t \in \mathcal{X}$ defined for $t = 1, ..., T+1$ with $x_1$ known. 
Our decision variables are $u_t \in \mathcal{U}$ for $t = 1, ..., T$ and each $u_t$ is
chosen as a function of the current state, $u_t = \mu_t(x_t)$.
(We use the same symbol as the volumes
traded at time $t$ since in the following they coincide.) 
The randomness of the system is modeled by a series of IID random 
variables $w_t \in \mathcal{W}$, for $t = 1, ..., T$.
The dynamics is described by a series of functions
\[
x_{t+1} = f_t(x_t, u_t, w_t),
\]
at every stage we incur the cost
\[
g_t(x_t, u_t, w_t),
\]
and at the end of the decision process we have a final cost
\[
g_{T+1}(x_{T+1}).
\]
Our objective is to minimize
\[
J = \expect \left[\sum_{t=1}^{T} g_t(x_t, u_t, w_t) + g_{T+1}(x_{T+1})\right].
\]
We solve the problem by \emph{backward induction},
defining the \emph{cost-to-go} function $v_t$ at each time step $t$
\BEQ
	\label{eq:bellman_1}
	v_t(x) = \min_u \expect [g_t(x, u, w_t) + v_{t+1}(f_t(x, u, w_t))], \quad t = 1,\ldots,T.
\EEQ
This recursion is known as \emph{Bellman} equation. The final condition is fixed by
\[
v_{T+1}(\cdot) = g_{T+1}(\cdot).
\]
It follows that the optimal action at time $t$ is given by the solution
\BEQ
	\label{eq:bellman_2}
	u_t = \argmin_u \expect [g_t(x_t, u, w_t) + v_{t+1}(f_t(x_t, u, w_t))].
\EEQ
In general, these equations are not solvable since the iteration that defines
the functions $v_t$ requires an amount of computation exponential in the
dimension of the state space, action space, and number of time steps
(\emph{curse of dimensionality}).
However some special forms of this problem have closed form solutions. We see
one in the following section.

\subsection{Linear-quadratic stochastic control}
\label{subsec:lin_quad_reg}
Whenever the dynamics functions $f_t$ are stochastic affine and the cost
functions are stochastic quadratic, the problem of \S\ref{subsec:dyn_prog}
has an analytic solution \cite{ee365}. 
We call this \emph{Linear-Quadratic Stochastic Control} (LQSC). 
We define the state space $\mathcal{X} = \reals^n$, the action space $\mathcal{U} = \reals^m$ 
for some $n,m > 0$. 
The disturbances are independent with known distributions and belong to a general set $\mathcal{W}$.
For $t = 1, \ldots, T$ the system dynamics is described by 
\[
x_{t+1} = f_t(x_t, u_t, w_t) = A_t(w_t) x_t + B_t(w_t) u_t + c_t(w_t), \  \  t = 1, ..., T
\]
with matrix functions $A_t(\cdot): \mathcal{W} \to \reals^{n\times n}$, $B_t(\cdot): \mathcal{W} \to \reals^{n\times m}$,
and $c_t(\cdot): \mathcal{W} \to \reals^n$.
The stage costs are
\[
g_t(x_t, u_t, w_t) = x_t^T Q_t(w_t)x_t + q_t(w_t)^T x_t + u_t^T R_t(w_t)u_t + r_t(w_t)^Tu_t 
\]
with matrix functions $Q_t(\cdot): \mathcal{W} \to \reals^{n\times n}$, 
$q_t(\cdot): \mathcal{W} \to \reals^n$,
$R_t(\cdot): \mathcal{W} \to \reals^{m\times m}$, and
$r_t(\cdot): \mathcal{W} \to \reals^m$.
The final cost is a quadratic function of the final state
\[
g_{T+1}(x_{T+1}) = x_{T+1}^T Q_{T+1}x_{T+1} + q_{T+1}^T x_{T+1}.
\]
The main result of the theory on linear-quadratic problems \cite{bertsekas1995dynamic}
 is that the optimal policy $\mu_t(x_t)$ is a simple affine function of the problem parameters and can be obtained analytically
\BEQ
\label{eq:riccati_1}
\mu_t(x_t) = K_t x_t + l_t, \  \  t = 0, ..., T-1,
\EEQ
where $K_t \in \reals^{m\times n}$ and $l_t \in \reals^{m}$ depend on the problem parameters. 
In addition, the cost-to-go function is a quadratic function of the state
\BEQ
\label{eq:riccati_2}
v_t(x_t) = x_t^T D_t x_t + d_t^T x_t + b_t
\EEQ
where $D_t \in \reals^{n \times n}$, $d_t \in \reals^n$, and $b_t \in \reals$ 
for $t = 1, \ldots, T$.  
We derive these results solving the Bellman equations \eqref{eq:bellman_1} by backward induction.
These are known as \emph{Riccati equations}, reported in Appendix \ref{appendix:subsec_riccati_eqs_LQSC}.

\subsection{Conditionally dependent disturbances}
\label{subsec:cond_dep_dist}
We now consider the case in which the disturbances are \emph{not} independent,
and we can't apply the Bellman iteration of \S\ref{subsec:dyn_prog}. 
Specifically, we assume that the disturbances have a joint distribution
described by a density function
\[
f_w(\cdot) : \mathcal{W} \times \cdots \times \mathcal{W} \to [0,1].
\]
One approach to solve this problem is to \emph{augment}
the state $x_t$, by including the disturbances observed up to time $t$.
This causes the computational complexity of the solution to grow exponentially
with the increased dimensionality (curse of dimensionality).
Some \emph{approximate dynamic programming} techniques can be used to solve
the augmented problem \cite{bertsekas1995dynamic} \cite{powell2007approximate}.
We take instead the approximate approach developed in \cite{skaf2010shrinking}, called
\emph{shrinking-horizon dynamic programming} (SHDP),
which performs reasonably well in practice and
leads to a tractable solution.
(It can be seen as an extension of \emph{model predictive control},
known to perform well in a variety of scenarios \cite{bemporad2006model} \cite{kwon2006receding}
\cite{mattingley2011receding} \cite{boyd2013performance}).

We now summarize the approach. Assume we know the density of the future disturbances $w_{t}, \ldots, w_T$
conditioned on the observed ones 
\[
f_{w|{t}}(w_{t}, \ldots, w_T) : \mathcal{W} \times \cdots \times \mathcal{W} \to [0,1].
\]
(If $t=1$ this is the unconditional density.) 
We derive the marginal density of 
each future disturbance, by integrating over all others,
\[
\hat f_{w_{t}|t}(w_{t}), \ldots, \hat f_{w_{T}|t}(w_{T}).
\]
We use the product of these marginals
to approximate the density of the future disturbances, so they all are independent.
We then compute the cost-to-go functions with backwards induction
using the Bellman equations \eqref{eq:bellman_1} and \eqref{eq:bellman_2},
where the expectations over each disturbance $w_\tau$
are taken on the conditional marginal density
$\hat f_{w_{\tau}|t}$.
The equations \eqref{eq:bellman_1} for the cost-to-go function become
(note the subscript ${\cdot|t}$)
\BEQ
\label{eq:bellman_shdp}
	v_{\tau|t}(x) = \min_{u} \expect_{ \hat f_{w_{\tau}|t}} [g_\tau(x, u, w_\tau) + v_{\tau+1|t}(f_\tau(x, u, w_\tau))],
\EEQ
for all times $\tau = t, \ldots, T$, with the usual final condition. 
Similarly, the equations \eqref{eq:bellman_2} for the optimal action become
\BEQ
\label{eq:bellman_solution_shdp}
u_t = \argmin_u \expect_{ \hat f_{w_{t}|t}} [g_t(x_t, u, w_t) + v_{t+1|t}(f_t(x_t, u, w_t))]
\EEQ
for all times $\tau = t, \ldots, T$.
We \emph{only} use the solution $u_t$ at time $t$.
In fact when we proceed to the next time step $t+1$ we rebuild the whole sequence 
of cost-to-go functions $v_{t+1|t+1}(x), \ldots, v_{T|t+1}(x)$ using the updated
marginal conditional densities 
and then solve (\ref{eq:bellman_solution_shdp}) to get $u_{t+1}$. 
With this framework we can solve the VWAP problem we developed in \S\ref{sec:probl_formulation}.

\subsection{VWAP problem as LQSC}
\label{subsec:vwap_as_lqc}
We now formulate the problem described in \S\ref{sec:probl_formulation} 
in the framework of \S\ref{subsec:lin_quad_reg}.
For $t= 1, \ldots, T+1$ we define the state as:
\BEQ
\label{eq:def_state}
x_t =  \left( 
\begin{array}{c}
\sum_{\tau=1}^{t-1} u_\tau \\
\sum_{\tau=1}^{t-1} m_\tau
 \end{array}
\right),
\EEQ
so that $x_1 = (0,0)$.
The action is $u_t$, the volume we trade during interval $t$, as defined 
in \S\ref{sec:probl_formulation}.

The disturbance is
\BEQ
\label{eq:def_disturbance}
w_t = \left( 
\begin{array}{c}
m_{t} \\
V
 \end{array}
\right)
\EEQ
where the second element is the total market volume
$ V = \sum_{t=1}^T m_t$.
With this definition the disturbances are not conditionally independent. 
In \S\ref{subsec:vwap_lqc_solution} we study their joint and marginal distributions.
We note that $V$, the second element of each $w_t$, is not observed after time $t$. (The theory we developed so far does not require the disturbances $w_t$ to be observed,
the Bellman equations only need expected values of functions of $w_t$.)
For $t=1, ..., T$ the state transition consists in
\[
x_{t+1} = x_t +  \left( 
\begin{array}{c}
 {u_{t}}\\
{m_{t}}
 \end{array}
\right).
\] 
So that the dynamics matrices are
\BEAS
A_t(w_t) &=& 
\left( 
\begin{array}{cc}
1&0 \\
0 &1 \\
 \end{array}
\right) \equiv I,\\
B_t(w_t) &=& 
\left( 
\begin{array}{c}
1 \\
0 \\
 \end{array}
\right) \equiv e_1,\\
c_t(w_t) &=& 
\left( 
\begin{array}{c}
0 \\
m_t\\
 \end{array}
\right).
\EEAS
The objective funtion \eqref{eq:objective_func_normalized} can
be written as 
$
\expect_m \sum_{t=1}^T g_t(x_t, u_t, w_t) 
$
where each stage cost is given by
\[
g_t(x_t, u_t, w_t) = 
\frac{ {s}_{t}}{2}\left(\alpha\frac{u_t^2}{C m_{t}} - \frac{u_t}{C}\right) +
\lambda { \sigma}_{t}^2 x_t^T \left( 
\begin{array}{cc}
\frac{1}{C^2} & -\frac{1}{CV} \\
-\frac{1}{CV} & \frac{1}{V^2} \\
 \end{array}
\right) x_t.
\]
The quadratic cost function terms are thus 
\BEAS
Q_t(w_t) &=& \lambda  \sigma_{t}^2 \left( 
\begin{array}{cc}
\frac{1}{C^2} & -{1}/{CV} \\
-{1}/{CV} & {1}/{V^2} \\
 \end{array}
\right)\\
q_t(w_t) &=& 0 \\
R_t(w_t) &=& \frac{\alpha  s_t}{2Cm_t}\\
r_t(w_t) &=& -\frac{ s_t}{2C} 
\EEAS
for $t = 1, \ldots, T$.
The constraint that the total executed volume is equal to $C$
imposes the last action
\[
u_T = \mu_{T}(x_{T}) = C - \sum_{t=1}^{T-1} u_t, \equiv K_T x_t + l_t
\]
with
\BEAS
K_T &=& 
-e_1^T\\
l_T &=& C.
\EEAS
This in turn fixes the value function at time $T$
\BEQ
\label{eq:final_value_func}
v_T(x_T) = \expect g_T(x_T, K_T x_t + l_t, w_t),
\EEQ
so we can treat $x_T$ as our final state and only consider 
the problem of choosing actions up to $u_{T-1}$. 
We are left with the constraint $u_t \geq 0$ for $t = 1,\ldots, T$.
Unfortunately this can not be enforced in the LQSC formalism.
We instead take the \emph{approximate dynamic programming} approach of \cite{keshavarz2014quadratic}. 
We allow $u_t$ to get negative sign and then project 
it on the set of feasible solutions. For every $t = 1, \ldots, T$ we compute
\[
 \max(u_t, 0)
\] 
and use it, instead of $u_t$, for our trading schedule.
This completes the formulation of our optimization problem 
into the linear-quadratic stochastic control framework.
We now focus on its solution, using the approximate approach of \S\ref{subsec:cond_dep_dist}.

\subsection{Solution in SHDP}
\label{subsec:vwap_lqc_solution}
We provide an approximate solution of the problem defined in \S\ref{subsec:vwap_as_lqc}
using the framework of shinking-horizon dynamic programming 
(summarized in \S\ref{subsec:cond_dep_dist}).
Consider a fixed time $t = 1, \ldots, T-1$.
We note that (unlike the assumption of \cite{skaf2010shrinking})
we do not observe the sequence of disturbances $w_1, \ldots, w_{t-1}$,
because the total volume $V$ is not known until the end of the day.
We only observe the sequence of market volumes $m_1, \ldots, m_{t-1}$.

If $f_m (m_1, \ldots, m_T)$ is the joint distribution of the market volumes,
then the joint distribution of the disturbances is
\[
f_w(w_1, \ldots, w_t) = f_m (e_1^Tw_1, \ldots, e_1^Tw_T) \times \mathbf{1}_{\{e_2^Tw_1 = V\}} \times \cdots \times
 \mathbf{1}_{\{e_2^Tw_T = V\}} \times  \mathbf{1}_{\{V = \sum_{\tau=1}^T e_1^Tw_\tau\}}
\]
where $e_1 = (1,0)$, $e_2 = (0,1)$, and the function
$\mathbf{1}_{\{\cdot\}}$ has value 1 when the condition is true and 0 otherwise.
We assume that our market volumes model also provides the  
 conditional density $f_{m|t}(m_{t}, \ldots, m_T)$
 of $m_{t}, \ldots, m_T$ given $m_1, \ldots, m_{t-1}$.
The conditional distribution of $V$ given $m_1, \ldots, m_{t-1}$ is 
\[
f_{V|t}(V) = \int\cdots\int f_{m|t}(m_{t}, \ldots, m_T) \mathbf{1}_{\{V = \sum_{\tau=1}^T m_\tau \}} dm_{t}\cdots dm_T
\]
(where the first $t-1$ market volumes are constants and the others are integration variables).
Let the marginal densities be
\[
\hat f_{m_{t}|t}(m_{t}), \ldots, \hat f_{m_{T}|t}(m_{T}).
\]
The marginal conditional densities of the disturbances are thus
\BEQ
\hat f_{w_\tau|t}(\cdot) = \hat f_{m_{\tau}|t}(\cdot) \times f_{V|t}(\cdot)
\EEQ
for $\tau = t, \ldots, T$. 

We use these to apply
the machinery of \S\ref{subsec:cond_dep_dist}, 
solve the Bellman equations and obtain the suboptimal SHDP policy at time $t$.
We compute the whole sequence of cost-to-go functions and policies 
at times $\tau = t, \ldots, T$. 
The cost-to-go functions are 
\BEQ
\label{eq:riccati_2_SHDP}
v_{\tau|t}(x_\tau) = x_\tau^T D_{\tau|t} x_\tau + d_{\tau|t} x_\tau + b_{\tau|t}
\EEQ
for $\tau = t, \ldots, T-1$.
The only difference with equation \eqref{eq:riccati_2} is the condition $|t$ in the subscript,
because expected values are taken over the marginal conditional densities $\hat f_{w_\tau|t}(\cdot)$.
Similarly, the policies are 
\BEQ
\label{eq:riccati_1_SHDP}
\mu_{\tau|t}(x_\tau) = K_{\tau|t} x_\tau + l_{\tau|t}
\EEQ
for $\tau = t, \ldots, T-1$. 
We report the equations for this recursion in Appendix \ref{appendix:shdp_solution_formulas_base}.
At every time step $t$ we compute the whole sequence of cost-to-go and policies,
in order to get the optimal action 
\BEQ
\label{eq:def_dynamic_solution_SHDP}
u_t^\star = \mu_{t|t}(x_t) = K_{\tau|t} x_\tau + l_{\tau|t}.
\EEQ
We then move to the next time step and repeat the whole process.
If we are not interested in computing the cost-to-go $v_{t|t}(x_t)$  
the equations simplify somewhat
(we disregard large part of the recursion and only compute what we need).
We develop these simplified formulas in Appendix \ref{appendix:shdp_solution_formulas}.

\section{Empirical results}
\label{sec:empirical_results}
We study the performance of the \emph{static solution} of \S\ref{sec:static_solution}
versus the \emph{dynamic solution} of \S\ref{sec:dynamic_sol}
by simulating execution or stock orders,
 using real NYSE market price and volume data. 
We describe in \S\ref{subsec:data} the dataset and how we process it. 
The dynamic solution requires a model for the joint
distribution of market volumes, here we use a simple model, explained
in \S\ref{subsec:market_volume_model}. 
(We expect that a more sophisticated model for market volumes
would improve the solution performance significantly.)
In \S\ref{subsec:rolling_est} we describe the ``rolling testing'' framework in 
which we operate. 
Our procedure is made up of two parts: 
the historical estimation of model parameters, explained in \S\ref{subsec:historical_estim},
 and the actual simulation of order execution, in \S\ref{subsec:testing_order_exec}.
Finally in \S\ref{subsec:aggregate_results} we show our aggregate results. 

\subsection{Data}
\label{subsec:data}
\begin{figure}[htb!]
\centerline{\includegraphics[width=21cm]{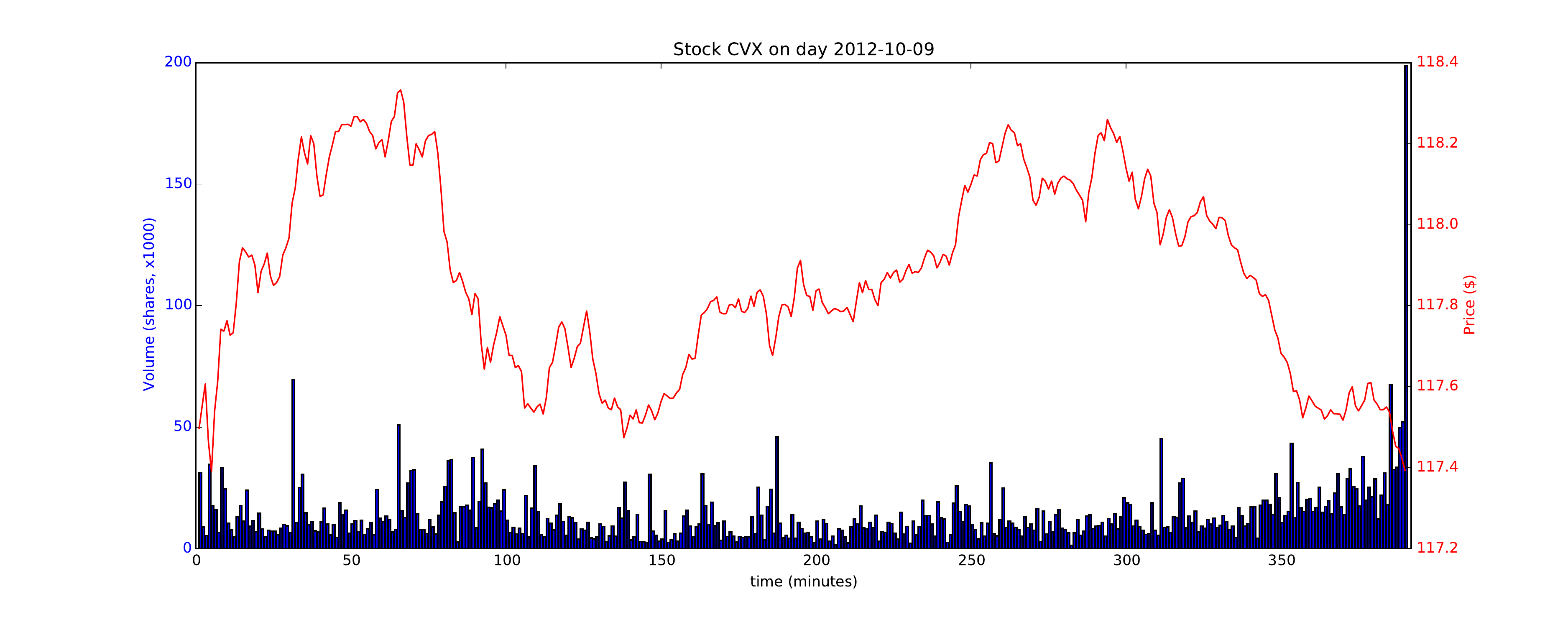}}
\caption{Example of a trading day. The blue bars are the market volumes traded every minute (in number of shares) and the red line is the period price $p_t$.}
\label{VolumePriceRealData}
\end{figure}

We simulate execution on data from the NYSE stock market.
Specifically, we use the $K=30$ different stocks which make up
the Dow Jones Industrial Average (DJIA), on $N = 60$ market days corresponding to the
last quarter of 2012, from September 24 to December 20 
(we do not consider the last days of December because 
the market was either closed or had reduced trading hours). 
The 30 symbols in that quarter are: MMM, AXP, T, BA, CAT, 
CVX, CSCO, KO, DD, XOM, GE, HD, INTC, IBM, JNJ, JPM, MCD, 
MRK, MSFT, PFE, PG, TRV, UNH, UTX, VZ, WMT, DIS, AA, BAC, HPQ.
We use raw Trade and Quotes (TAQ) data from Wharton Research Data Services
 (WRDS) \cite{TAQ_WRDS}.  
We processe the raw data to obtain daily series of market volumes $m_t \in \integers_{+}$ 
and average period price $p \in \reals_{++}$, for $t = 1, \ldots, T$ where $T = 390$, 
so that each interval is one minute long. 
We clean the raw data by filtering out trades meeting any of the following conditions:
\begin{itemize}
\item \emph{correction code} greater than 1, trade data incorrect;
\item \emph{sales condition} ``4'', ``@4'', ``C4'', ``N4'', ``R4'', 
  \emph{derivatively priced}, \ie, the trade 
 was executed over-the-counter (or in an external facility like a Dark Pool);
\item \emph{sales condition} ``T'' or ``U'',  \emph{extended hours} 
trades (before or after the official market hours);
\item \emph{sales condition} ``V'', \emph{stock option} trades (which are also executed over-the-counter);
\item  \emph{sales condition} ``Q'', ``O'', ``M'', ``6'', 
 \emph{opening trades} and \emph{closing trades}
(the opening and closing auctions).
\end{itemize}
In other words we focus exclusively on the continuous trading activity
without considering market opening and closing nor any over-the-counter trade.
In Figure \ref{VolumePriceRealData} we plot an example of market volumes and prices.

\subsection{Market volumes model}
\label{subsec:market_volume_model}
We have so far assumed that the distribution of market volumes
\[
f_m(m_1, \ldots, m_T)
\]
is known from the start of the day. 
In reality a broker has a parametric family of distributions 
and each day (or less often) selects the parameters for the distribution  
with some statistical procedure. For simplicity, 
we assume such procedure is based on historical data.
We found few works in the literature concerned with 
intraday market volumes modeling (\cite{bialkowski2008improving}). 
We thus develop our own market volume model. 
This is composed of a parametric family of market volume distributions and
an \emph{ad hoc} procedure to choose the parameters with historical data.

For each stock we model the vector of market volumes as a multivariate log-normal.
If the superscript $(k)$ refers to the stock $k$
(\ie,
$m_t^{(k)}$ is the market volume for stock $k$ in interval $t$),
we have 
\BEA
  \label{eqDefVolumeModel}
  f_{m^{(k)}}(m^{(k)}_1, \ldots, m^{(k)}_T) \sim  \ln\mathcal{N}(\mu + \ones b^{(k)}, \Sigma)
\EEA
where $b^{(k)} \in \reals$ is a constant that depends 
on the stock $k$ (each stock has a different typical daily volume),
$\mu \in \reals^T$ is an average ``volume profile'' (normalized so that $\ones^T \mu = 0$)
and $\Sigma \in \symm_{++}^T$ is a covariance matrix.
The volume process thus separates
into a per-stock deterministic component, modeled by the constant $b^{(k)}$,
and a stochastic component with the same distribution for all stocks, 
modeled as a multivariate log-normal.
We report in Appendix \ref{appendix:volume_model} the \emph{ad hoc} procedure
we use to estimate the parameters of this volume model on historical 
data and the formulas for the conditional 
expectations $\expect_t \left[{1}/{V}\right]$, $\expect_t [m_\tau]$, $\expect_t \left[{1}/{m_\tau}\right]$
for $\tau = t, \ldots, T$ (which we need for the solution \eqref{eq:def_dynamic_solution_SHDP}).
The procedure for estimating the volume model on past data 
requires us to provide a parameter, which we estimate with
cross-validation on the initial section of the data. The details
are explained in Appendix \ref{appendix:volume_model}.

\subsection{Rolling testing}
\label{subsec:rolling_est}
We organize our simulations according to a ``rolling testing'' or ``moving window'' procedure:
for every day used to simulate order execution we estimate the various 
parameters on data from a ``window'' covering the preceding $W>0$ days.
(It is commonly assumed that the most recent historical data 
are most relevant for model calibration since the systems underlying 
the observed phenomena change over time). 
We thus simulate execution on each day $i = W + 1, \ldots, N$ 
using data from the days $i - W, \ldots, i - 1$ for historical estimation.

In this way every time we test a VWAP solution algorithm, we use model parameters
 calibrated on historical data exclusively. 
 In other words  
 the performance of our models are estimated \emph{out-of-sample}.
In addition since all the order simulations use the same amount of historical data
for calibration it is fair to compare them.

We fix the window lenght of the historical estimation to $W = 20$, 
corresponding roughly to one month. 
We set aside the first $W_{CV} = 10$ simulation days 
 for cross-validating
a feature of the volume model, as explained in Appendix \ref{subsec:appendix:volume_cross_val}.
In \figref{fig:rolling_testing_schematic}
we describe the procedure. In the next two sections we explain 
how we perform the estimation of model parameters and simulation of orders execution.
\begin{figure}[htb]
\centering 
\begin{tikzpicture}
\draw (0,0) -- (12,0);
\draw [dashed, ->] (12,0) -- (15,0);
\filldraw [fill=black!20,draw=black!20](0,0) rectangle (6,.2);
\filldraw [fill=green!40,draw=green!40](6,0) rectangle (6.3,.2) 
  node[black, midway, yshift=.5cm] {\footnotesize order simul. (cross. val.)};
\draw[->] (6.15,.45) -- (6.15,.25);
\draw[step=.3cm] (0,0) grid (12,.2);
\draw [thick,decorate,decoration={brace,amplitude=8pt,mirror},yshift=-2pt](0, 0) -- (6,0) node[black,midway,yshift=-0.5cm] {\footnotesize $W=20$ days};

\draw [yshift=-1.5cm](0,0) -- (12,0);
\draw [dashed, ->,yshift=-1.5cm] (12,0) -- (15,0);
\filldraw [fill=black!20,draw=black!20,yshift=-1.5cm](.3,0) rectangle (6.3,.2);
\filldraw [fill=green!40,draw=green!40,yshift=-1.5cm](6.3,0) rectangle (6.6,.2)
  node[black, midway, yshift=.5cm] {\footnotesize order simul. (cross. val.)};
\draw[->,yshift=-1.5cm] (6.45,.45) -- (6.45,.25);
\draw[step=.3cm,yshift=-1.5cm] (0,0) grid (12,.2);
\draw [thick,decorate,decoration={brace,amplitude=8pt,mirror},yshift=-1.5cm-2pt](.3, 0) -- (6.3,0) node[black,midway,yshift=-0.5cm] {\footnotesize $W=20$ days};

\draw (7.5,-2.72)  node {.} ;
\draw (7.5,-2.6)  node {.} ;
\draw (7.5,-2.48)  node {.} ;

\draw [yshift=-4cm](0,0) -- (12,0);
\draw [dashed, ->,yshift=-4cm] (12,0) -- (15,0);
\filldraw [fill=black!20,draw=black!20,yshift=-4cm](3,0) rectangle (9,.2);
\filldraw [fill=green!40,draw=green!40,yshift=-4cm](6,.12) rectangle (9,.2);
\draw [thick,green!0!black,decorate,decoration={brace,amplitude=5pt},yshift=-4cm+2pt](6, .2) -- (9,.2) node[green!0!black,midway,yshift=0.4cm] {\footnotesize $W_{CV} = 10$ days};
\filldraw [fill=red!40,draw=red!40,yshift=-4cm](9.0,0) rectangle (9.3,.2)
  node[black, midway, yshift=-.6cm] {\footnotesize order simul. };
\draw[->,yshift=-4cm] (9.15,-.3) -- (9.15,-.05);
\draw[step=.3cm,yshift=-4cm] (0,0) grid (12,.2);
\draw [thick,decorate,decoration={brace,amplitude=8pt,mirror},yshift=-4cm-2pt](3, 0) -- (9,0) node[black,midway,yshift=-0.5cm] {\footnotesize $W=20$ days};

\draw [yshift=-5.5cm](0,0) -- (12,0);
\draw [dashed, ->,yshift=-5.5cm] (12,0) -- (15,0);
\filldraw [fill=black!20,draw=black!20,yshift=-5.5cm](3.3,0) rectangle (9.3,.2);
\filldraw [fill=green!40,draw=green!40,yshift=-5.5cm](6,.12) rectangle (9,.2);
\filldraw [fill=red!40,draw=red!40,yshift=-5.5cm](9.3,0) rectangle (9.6,.2)
  node[black, midway, yshift=-.6cm] {\footnotesize order simul. };
\draw[->,yshift=-5.5cm] (9.45,-.3) -- (9.45,-.05);
\draw[step=.3cm,yshift=-5.5cm] (0,0) grid (12,.2);
\draw [thick,decorate,decoration={brace,amplitude=8pt,mirror},yshift=-5.5cm-2pt](3.3, 0) -- (9.3,0) node[black,midway,yshift=-0.5cm] {\footnotesize $W=20$ days};

\draw (7.5,-6.72)  node {.} ;
\draw (7.5,-6.6)  node {.} ;
\draw (7.5,-6.48)  node {.} ;
\end{tikzpicture}
\caption{Description of the rolling testing procedure. We iterate over the dataset, 
simulating execution on any day $i = W + 1, \ldots, N$
and estimating the model parameters on the preceding $W=20$ days.
 The first $W_{CV} = 10$ days used to simulate orders
  are reserved for cross validation (as explained in Appendix \ref{subsec:appendix:volume_cross_val}). The aggregate results from the remaining $W + W_{CV} + 1, \ldots, N$ days
(30 days in total) are presented in \S\ref{subsec:aggregate_results}.}
\label{fig:rolling_testing_schematic}
\end{figure}
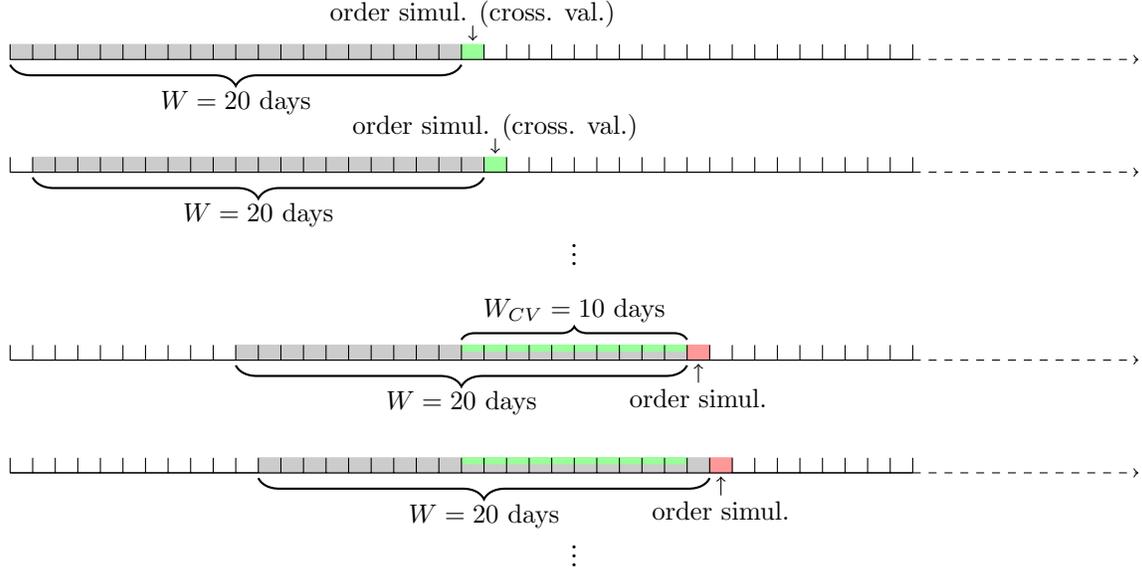

\subsection{Models estimation}
\label{subsec:historical_estim}
\begin{figure}
	\centerline{\includegraphics[width=20cm]{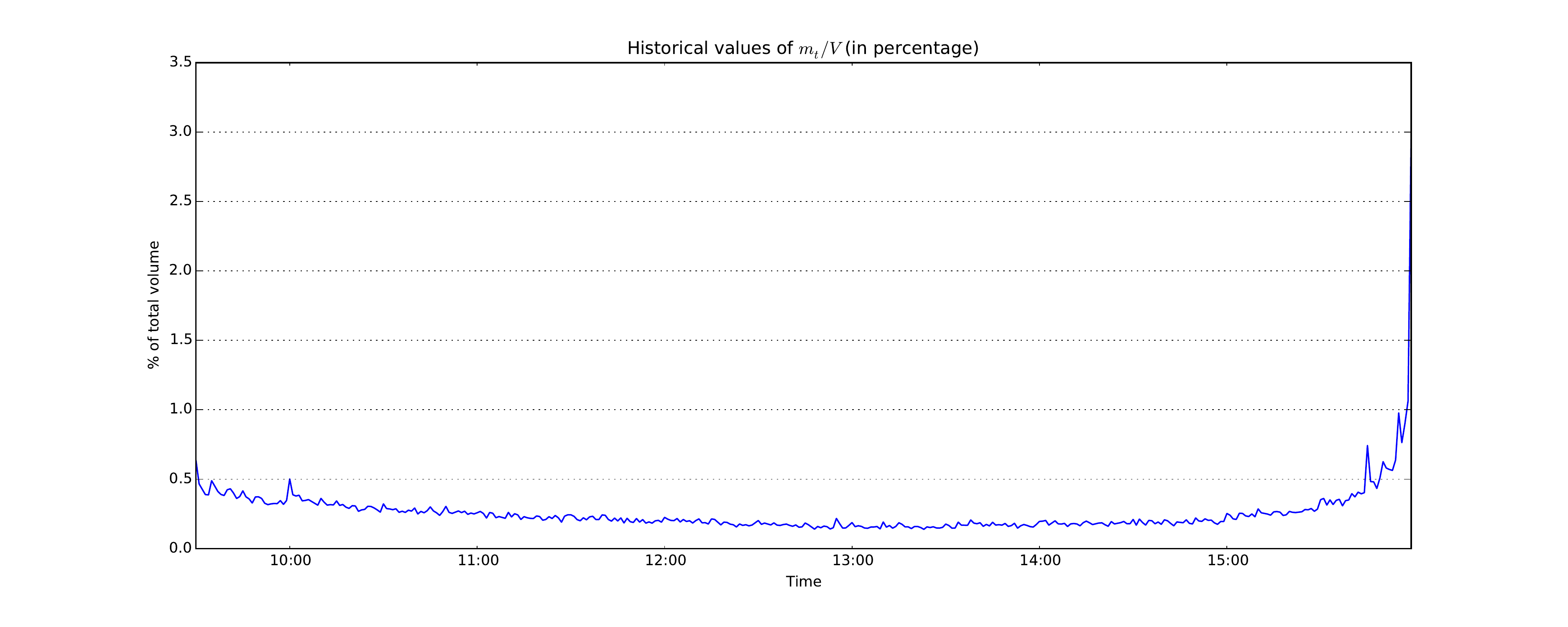}}
	\caption{Estimated values of $\expect \left[\frac{m_t}{V}\right]$ using
	the first $W = 20$ days of our dataset, shown in percentage points.}
	\label{fig:volume_profile_estimation}
\end{figure}

\begin{figure}
	\centerline{\includegraphics[width=20cm]{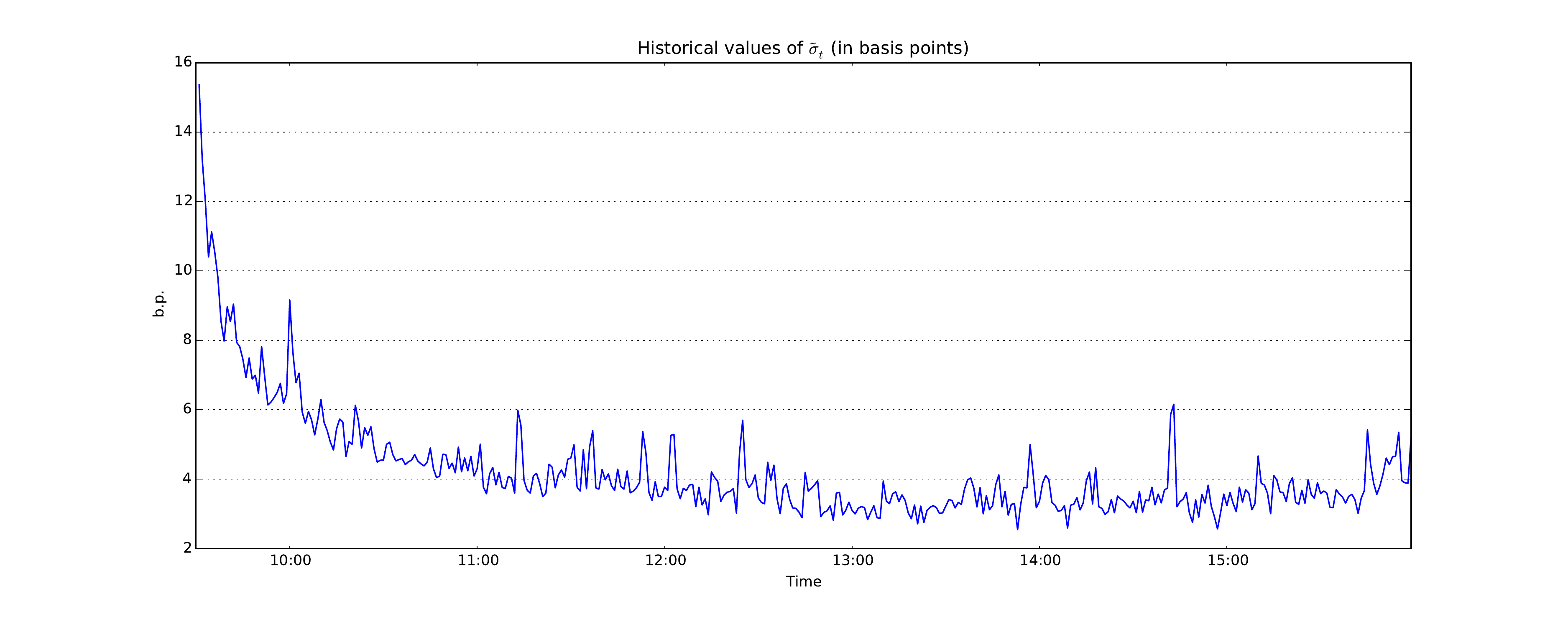}}
	\caption{Estimated values of the period volatilities, $\hat \sigma_t$ using
	the first $W = 20$ days of our dataset. For each period of one minute these are the estimated
	standard deviation of the price increments, shown in basis points (one basis point is $0.0001$).}
	\label{fig:sigmas_estimation}
\end{figure}

We describe the estimation, on historical data, of the parameters 
of all relevant models for our solution algorithms. 
We append the superscript $(i,k)$ to any
quantity that refers to market day $i$ and stock $k$.
We start by the market volumes per interval as a fraction of the total daily volume 
(which we need for \eqref{eq:solutiona_static_no_spread}). We use the sample average
\[
\expect \left[\frac{m_t}{V}\right] \simeq \frac{\sum_{j = i - W}^{i-1} \sum_{k=1}^K {m_t^{(j,k)}}/{V^{(j,k)}}}{W K}
\]
for every $t = 1, \ldots, T$.  An example of this estimation (on the first $W = 20$ days of the dataset)
is shown in \figref{fig:volume_profile_estimation}.
The dynamic solution \eqref{eq:def_dynamic_solution_SHDP} requires an estimate of the
 volatilites $ \sigma_t$, we use the sample average of the squared price changes
\[
  \sigma_t^2 \simeq \frac{\sum_{j = i - W}^{i-1} \sum_{k=1}^K {\left((p^{(j,k)}_{t+1} - p^{(j,k)}_t)/p^{(j,k)}_t\right)}^2 }{W K}
\]
for every $t = 1, \ldots, T$. In \figref{fig:sigmas_estimation} we show an example of
 this estimation (on the first $W$ days of the dataset).
We then choose the volume distribution $f_m(m_1, \ldots, m_T)$ 
among the parametric family defined in \S\ref{subsec:market_volume_model} (using the \emph{ad hoc} procedure described in Appendix \ref{subsec:appendix:vol_estimation}). 
We estimate 
the expected daily volume for each stock
as the sample average 
\[
\expect[ V^{(i, k)}] \simeq \frac{\sum_{j = i - W}^{i-1} {V^{(j,k)}}}{W}
\]
for every $k = 1, \ldots, K$. We use this to choose the size of the simulated orders.

Finally, we consider the parameters $s_1, \ldots, s_T$, and $\alpha$ 
of the transaction cost model \eqref{eq:def:trans_cost_model}. 
We do not estimate them empirically since we would
 need additional data, market quotes for the spread 
and proprietary data of executed orders for $\alpha$ (confidential
for fiduciary reasons).
We instead set them to exogenous values, kept
constant across all stocks and days (to simplify comparison of execution costs).
We assume for simplicity that the fractional spread is constant in time
and equal to $2$ basis points, $s_1 = \cdots =  s_T = 2 \text{ b.p.}$ 
(one basis point is $0.0001$). That is reasonable for liquid stocks such 
as the ones from the DJIA.
We choose the parameter $\alpha$ following a rule-of-thumb of transaction costs:
trading one day's volume costs approximately on day's volatility \cite{kissell2003optimal}.
We estimate empirically over the first 20 days of the dataset the open-to-close
volatility for our stocks, equal to approximately $90$ basis points, and thus from
 equation \eqref{eq:def:trans_cost_model} we set $\alpha = 90$.

\subsection{Simulation of execution with VWAP solution algorithms}
\label{subsec:testing_order_exec}
For each day $i = W+1, \ldots, N$ and
each stock $k = 1, \ldots, K$
we simulate the execution of a trading order.
We fix the size of the order  
equal to $1\%$ of the expected daily volume for the given stock 
on the given day 
\[
C^{(i,k)} = \expect[ V^{(i, k)}]/100.
\]
Such orders are small enough to have negligible impact on 
the price of the stock \cite{bouchaud2009markets}, as we need
for \eqref{eq:def:trans_cost_model} to hold.

We repeat the simulation with different solution methods:
the static solution \eqref{eq:solutiona_static_no_spread}
and the dynamic solution \eqref{eq:def_dynamic_solution_SHDP}
with risk-aversion parameters $\lambda = 0, 1, 10, 100, 1000, \infty$.
We use the symbol $a$ to index the solution methods.
For each simulation we solve the appropriate set
of equations, setting all historically estimated parameters to the values
obtained with the procedures of \S\ref{subsec:historical_estim}. 
For each solution method we obtain a simulated 
trading schedule
\[
u_t^{(i,k,a)}, \quad t = 1, \ldots, T
\]
where the superscript $a$ indexes the solution methods.
We then compute the slippage incurred by the schedule using \eqref{eq:normalized_slippage}
\BEQ
\label{eq:def_simulated_slippage}
 S^{(i,k,a)} = \frac{\sum_{t=1}^T p^{(i,k)}_t u_t^{(i,k,a)} - C^{(i,k)}p^{(i,k)}_\text{VWAP}}{C^{(i,k)}p^{(i,k)}_\text{VWAP}}  + \sum_{t=1}^T \frac{s_t}{2}\left(\alpha \frac{{(u^{(i,k,a)}_t)}^2}{C^{(i,k)}m^{(i,k)}_t} - \frac{u^{(i,k,a)}_t}{C^{(i,k)}}\right).
\EEQ

Note that we are simulating the transaction costs.
Measuring them directly would require to actually execute $u_t^{(i,k,a)}$.
This test of transaction costs optimization has value
as a comparison between the static solution
 \eqref{eq:solutiona_static_no_spread}
 and the dynamic solution \eqref{eq:def_dynamic_solution_SHDP}.
 Our transaction costs model \eqref{eq:def:trans_cost_model} 
 is similar to the ones of other
works in the literature (\eg, \cite{frei2013optimal}) but involves
the market volumes $m_t$. 
The static solution only uses the market volumes distribution known before the market opens,
while the dynamic solution uses the SHDP procedure to incorporate real time information and
improve modeling of market volumes.
In the following we show that the dynamic solution achieves 
lower transaction costs than the static solution,
such gains are due to the better 
handlng of information on market volumes. 

In practice a broker would use a different model of transaction
costs, perhaps more complicated than ours. We think that
a good model should incorporate the market volumes $m_t$ as a
key variable \cite{bouchaud2009markets}. Our
test thus suggests that also in that setting the dynamic solution would 
perform better than the static solution.

\begin{figure}[htb!]
	\centerline{\includegraphics[width=19cm]{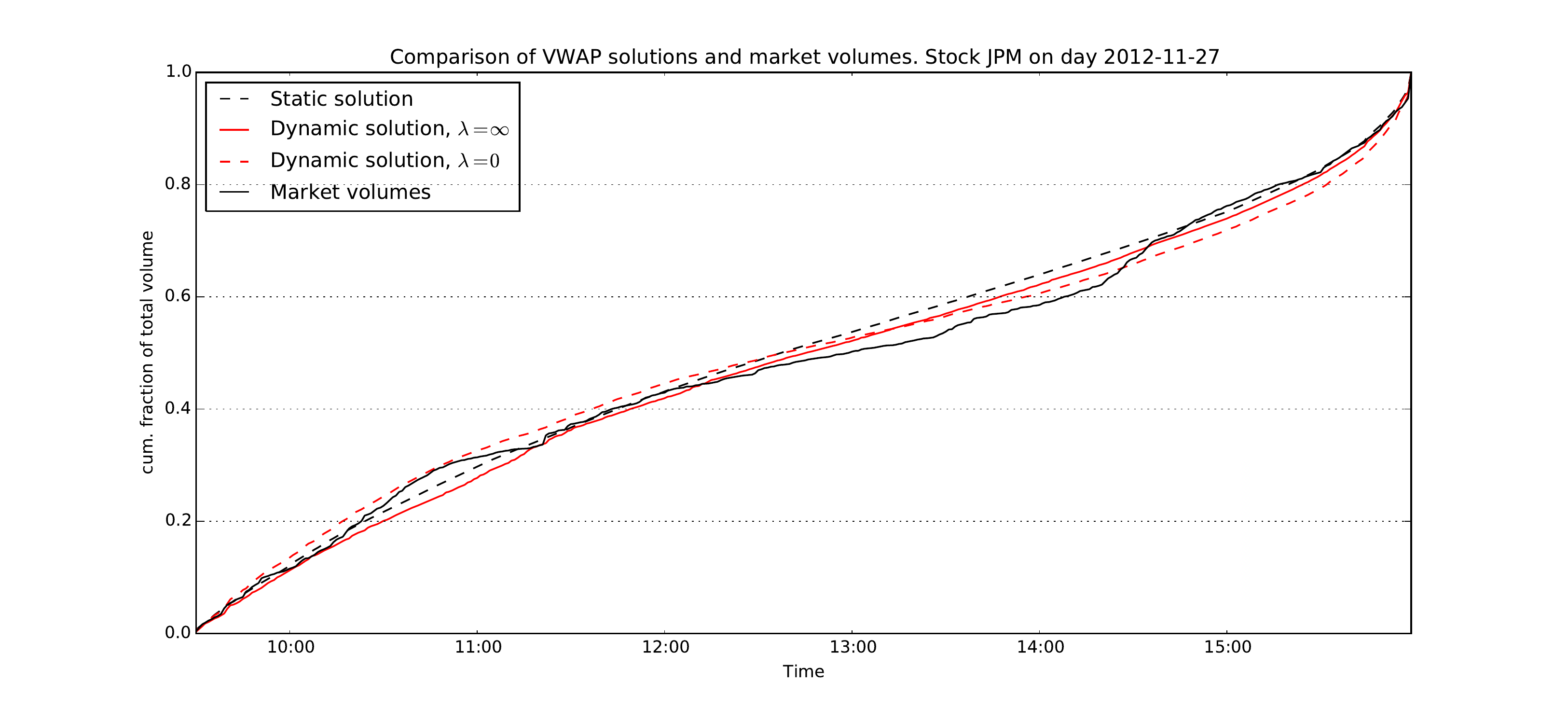}}
	\caption{Simulation of order execution on a sample market day. 
	We report all volume processes as cumulative fraction of their total. 
	At every time $\tau$
	we plot $\sum_{t=1}^\tau\frac{m_t}{V}$ for the market volumes $m_t$ 
	and $\sum_{t=1}^\tau\frac{u_t}{C}$ for the various solutions $u_t$.
  We only show the dynamic solution for $\lambda = 0$
  and $\lambda= \infty$ since for all other values of $\lambda$ the solution falls in between.
	}
	\label{fig:sample_day_solution}
\end{figure}

We show in \figref{fig:sample_day_solution} the result of the simulation
on a sample market day, using the static solution  \eqref{eq:solutiona_static_no_spread}
and the dynamic solution \eqref{eq:def_dynamic_solution_SHDP} for $\lambda = 0$ and $\infty$.
We also plot the market volumes $m_t^{(i,k)}$. 

\subsection{Aggregate results}
\label{subsec:aggregate_results}
\begin{figure}[h!]
  \centerline{\includegraphics[width=19cm]{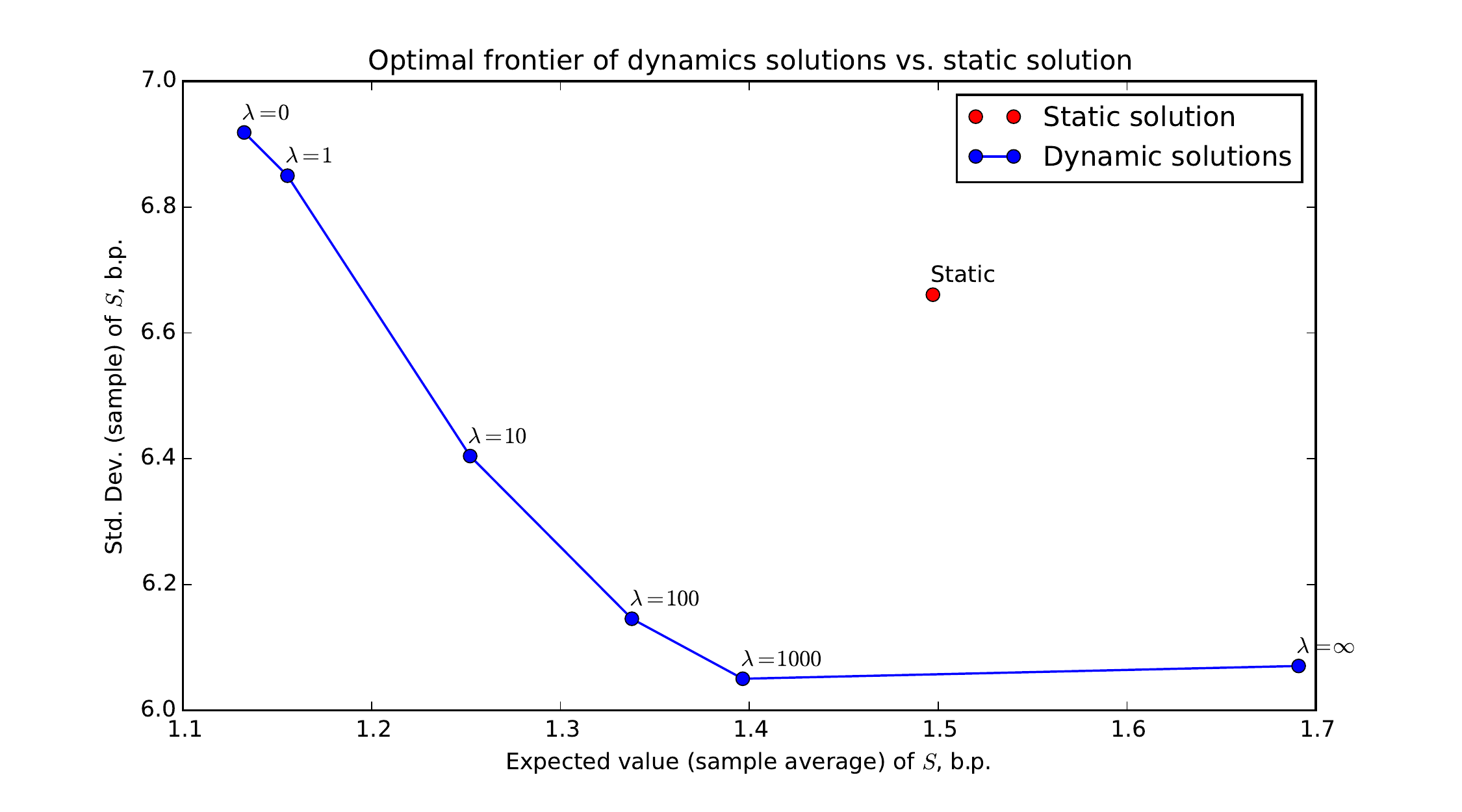}}
  \caption{Risk-reward plot of the aggregate results of our simulations on real market data. 
  Each dot represents one solution method, either the static solution \eqref{eq:solutiona_static_no_spread}
or the \emph{dynamic solution} \eqref{eq:def_dynamic_solution_SHDP}
with risk-aversion parameters $\lambda = 0, 1, 10, 100, 1000, \infty$. We 
show the sample average of the simulated slippages, which represents the 
execution costs, and the sample standard deviation, \ie, the
root mean square error (RMSE) of tracking the VWAP. The orders have size
equal to $1\%$ of the expected daily volume. The dynamic solution
improves over the static solution in both dimensions, we can choose
the preferred behaviour by fixing the risk-aversion parameter $\lambda$.}
  \label{fig:optimal frontier}
\end{figure}

We report the aggregate results from the simulation of VWAP execution on 
all the days reserved for orders simulation (minus the ones used for cross-validation).
For any day $i = W+W_{CV}+1, \ldots, N$, 
stock $k = 1, \ldots, K$, and solution method $a$ 
(either the static solution \eqref{eq:solutiona_static_no_spread}
or the dynamic solution \eqref{eq:def_dynamic_solution_SHDP} for various
values of $\lambda$) we obtain the 
 simulated slippage $S^{(i,k,a)}$ using \eqref{eq:def_simulated_slippage}.
Then, for each solution method $a$ we define the empirical expected
value of $S$ as
\[
\expect [S^{(a)}] = \frac{\sum_{i =  W+W_{CV}+1}^N \sum_{k=1}^K S^{(i,k,a)}}{(N- W -W_{CV})K}
\]
and the empirical variance
\[
\var (S^{(a)}) = \frac{\sum_{i =  W+W_{CV}+1}^N \sum_{k=1}^K {(S^{(i,k,a)})}^2 - 
\expect [S^{(a)}]^2}{(N- W -W_{CV})K -1}.
\]
In Figure \ref{fig:optimal frontier} we show the values of these on a risk-reward plot. 
(We show the square root of the variance
for simplicity, so that both axes are expressed in basis points). 
We observe that the dynamic solution improves over the static solution on both
VWAP tracking (variance of $S$) and transaction costs (expected value of $S$),
and we can select between the different behaviors by choosing different
values of $\lambda$.

We introduced in \S\ref{subsec:probl_objective} the approximation 
that the value of \eqref{eq:second_term_variance} is negligible when compared to 
\eqref{eq:first_term_variance}. The empirical results validate this. 
For the static solution
the empirical average value of \eqref{eq:first_term_variance} is $4.45e-07$
while \eqref{eq:second_term_variance} is $6.34e-09$, about $1\%$. 
For the dynamic solution with $\lambda = \infty$ the average value of 
\eqref{eq:second_term_variance} is $3.60e-07$ and
\eqref{eq:first_term_variance} is $1.92e-08$, about $5\%$.
For the dynamic solution with $\lambda = 0$ instead the average value of 
\eqref{eq:second_term_variance} is $4.76e-07$ and
\eqref{eq:first_term_variance} is $5.50e-09$, about $1\%$.
The dynamic solutions for other values of $\lambda$ sit 
in between. Thus the approximation is generally valid,
 becoming less tight for high values of $\lambda$. In fact in Figure \ref{fig:optimal frontier}
 we see that the empirical variance of $S$ for the dynamic solution with
 $\lambda = \infty$ is somewhat larger than the one with $\lambda = 10000$,
 probably because of the contribution of \eqref{eq:first_term_variance}.
 (We can interpret this as a \emph{bias-variance} tradeoff
 since by going from $\lambda=\infty$ to $\lambda = 10000$ we
 effectively introduce a regularization of the solution.)

\section{Conclusions}
\label{sec:conclusions}
We studied the problem of optimal execution under VWAP benchmark and developed
two broad families of solutions. 

The static solution of \S\ref{sec:static_solution},
although derived with similar assumptions to the classic \cite{konishi2002optimal}, is more 
flexible and can accommodate more sophisticated models (of bid-ask spread and volume)
 than the comparable static solutions in the literature.
 By formulating the problem as a quadratic program
 it is easy to add other convex constraints (see \cite{moallemi2012dynamic} for a good list)
 with a guaranteed straightforward fast solution \cite{boyd2009convex}. 

The dynamic solution of \S\ref{sec:dynamic_sol} is the biggest contribution
of this work. 
One one side, we manipulate the problem to fit it into the standard formalism
of linear-quadratic stochastic control. On the other, we model the uncertainty
on the total market volume (which is eschewed in all similar works we found in the
literature) in a principled way, building on a recent result in optimal control 
\cite{skaf2010shrinking}.

The empirical tests of \S\ref{sec:empirical_results} are based on simulations with real data
designed with good statistical practices (the rolling testing of \S\ref{subsec:rolling_est}
ensures that all results are obtained out-of-sample). We compare the performance of 
the static solution, standard in the trading industry, 
 to our dynamic solution. The dynamic solution is built around a model
for the joint distribution of market volumes,
we provide a simple one in \S\ref{subsec:market_volume_model} (along with \emph{ad hoc}
 procedures to use it).
This is supposed to be a proof-of-concept since
in practice a broker would have a more sophisticated market volume model, 
which would further improve performance of the dynamic solution.
Even with our model for market volumes
our dynamic solution improves the 
performance of the
static solution significantly. 
The result validates all the approximations involved
in the derivation of the dynamic solution and thus shows its value.

Our simulations quantify the improvements of our
dynamic solution over the standard static solution. On one side
we can reduce the RMSE of VWAP tracking by $10\%$.
This is highly significant and could improve with a more sophisticated market volume model. 
On the other we can lower the execution costs by around $25\%$.
In our test this corresponds to $\sim50\$$ of savings 
for an order of a million dollars
(the VWAP executions are worth billions of dollars each day). 

\newpage
\begin{appendices}

\section{Dynamic programming equations}
\subsection{Riccati equations for LQSC}
\label{appendix:subsec_riccati_eqs_LQSC}
We derive the recursive formulas for \eqref{eq:riccati_1} and \eqref{eq:riccati_2}. 
We know the final condition
\[
v_{T+1}(x_{T+1}) = g_{T+1}(x_{T+1}),
\]
so $D_{T+1} =  Q_{T+1}$, $d_{T+1} =  d_{T+1}$, and $b_{T+1} = 0$. Now for
the inductive step, assume $v_{t+1}(x_{t+1})$ is in the form of (\ref{eq:riccati_2})
with known $D_{t+1}$,  $d_{t+1}$, and $b_{t+1}$. Then the optimal action at time $t$ is,
according to \eqref{eq:bellman_2}, 
\[
u_t =
\argmin_u \Expect [g_t(x_t, u, w_t) + v_{t+1}(A_t(w_t) x_t + B_t(w_t) u + c_t(w_t))] = \\
 K_t x_t + l_t,
\]
with
\BEAS
K_t &=& - \frac{ \Expect [B_t(w_t)^T D_{t+1}A_t(w_t)] }{(\Expect R_t(w_t)  + \Expect[B_t(w_t)^T D_{t+1}B_t(w_t)]}\\
l_t &=& - \frac{ \Expect r_t  + 2\Expect [B_t(w_t)^T D_{t+1}c(w_t)]+ d_{t+1}^T\Expect B_t(w_t)}{2(\Expect R_t(w_t)  
+ \Expect[B_t(w_t)^T D_{t+1}B_t(w_t)]}.
\EEAS
It follows that the value function at time $t$ is 
also in the form of (\ref{eq:riccati_2}), and it has value
\begin{multline*}
v_t(x_t) = 
\Expect \left[g_t(x_t, K_t x_t + l_t, w_t) + v_{t+1}(A_t(w_t) x_t + B_t(w_t) ( K_t x_t + l_t) + c_t(w_t))  \right] = \\
x_t^T D_t x_t + d_t^T x_t + b_t
\end{multline*}
with
\BEAS
D_t &=& \expect Q_t(w_t) + K_t^T \Expect \left [R_t(w_t) + B_t(w_t)^T D_{t+1} B_t(w_t) \right] K_t + \\
&& \expect [A_t(w_t)^T D_{t+1}A_t(w_t)]+ K_t^T \expect [B_t(w_t)^T D_{t+1}A_t(w_t)] + \\
&&\expect [A_t(w_t)^T D_{t+1}B_t(w_t)] K_t \\
d_t &=& \expect q_t(w_t) + K^T_t \expect r_t(w_t)  + 2 \Expect K^T_t  R_t(w_t) l_t + \\
&&\expect [A_t(w_t) + B_t(w_t) K_t)^T(d_{t+1} + 2 D_{t+1} (B_t(w_t) l_t + \Expect c(w_t))]\\
b_t &=& b_{t+1} + \expect R_t(w_t) l_t^2 + \expect r_t(w_t) l_t  + \\
&&\expect [ (B_t(w_t) l_t +  c(w_t))^T D_{t+1} + d_{t+1}^T)(B_t(w_t) l_t +  c(w_t) )].
\EEAS
We thus completed the induction step, and so
 the value function is quadratic and the policy affine 
at every time step $t = 1, \ldots, T$. 
The recursion can be solved as long as we know the functional form of the problem parameters and 
the distribution of the disturbances $w_t$. 

\subsection{SHDP Solution}
\label{appendix:shdp_solution_formulas_base}
We derive the recursive formulas for \eqref{eq:riccati_2_SHDP} 
and \eqref{eq:riccati_1_SHDP}.
These are equivalent to the Riccati equations we derived
in Appendix \ref{appendix:subsec_riccati_eqs_LQSC}, but the 
expected values are taken over the marginal conditional densities $\hat f_{w_\tau|t}(\cdot)$.
We write $\expect_t$ to denote such expectation.
In addition, these equations are somewhat simpler since our problem has
$A_t(w_t) = I $, $B_t(w_t) = e_1$, $q_t = 0$, and $r_t(w_t) = r_t$ 
for $t = 1, \ldots, T$.
The final conditions are fixed by (\ref{eq:final_value_func}) 
\BEAS
D_{T|t} &=& \expect_t Q_T(w_T) +  e_1 \expect_t R_T(w_T) e_1^T\\
d_{T|t} &=& - r_T e_1^T - 2 C \expect_t R_T(w_T) e_1^T,\\
b_{T|t} &=& r_T C.
\EEAS 
And the recursive equations are
\BEAS
\label{eqnOptimalAction:const}
K_{\tau|t} &=& - \frac{e_1^T D_{\tau+1|t}}{\Expect_t R_\tau(w_\tau)  + e_1^T D_{\tau+1}e_1} \\
\label{eqnOptimalAction:lin}
l_{\tau|t} &=& - \frac{r_\tau + d_{\tau+1|t}^Te_1 + 2e_1^TD_{\tau+1|t}\Expect_t c(w_\tau) }{2(\Expect_t R_\tau(w_\tau) + e_1^T D_{\tau+1|t}e_1)}\\
D_{\tau|t} &=& \expect_t Q_\tau(w_\tau) + K^T_{\tau|t} \left( \Expect_t [R_\tau(w_\tau)] + e_1^T D_{\tau+1|t} e_1 \right)K_{\tau|t} +  \nonumber\\
&&D_{\tau+1|t} + K_{\tau|t}^T e_1^TD_{\tau+1|t} + D_{\tau+1|t} e_1 K_{\tau|t} =\nonumber\\
&& \expect_t Q_\tau(w_\tau) + D_{\tau+1|t} + K_{\tau|t}^T e_1^TD_{\tau+1|t} \label{eqnOptimalValue:quad}\\
d_{\tau|t} &=& K_{\tau|t}^T r_\tau + 2  K_{\tau|t}^T \Expect_t R_\tau(w_\tau) l_{\tau|t} + \nonumber \\
&&(I + e_1 K_{\tau|t})^T( d_{\tau+1|t} + 2 D_{\tau+1|t} (e_1 l_{\tau|t} + \Expect_t c(w_\tau))) = \nonumber\\
&& d_{\tau+1|t} + 2 D_{\tau+1|t} (e_1 l_{\tau|t} + \Expect_t c(w_\tau)) \label{eqnOptimalValue:lin}
\\
b_{\tau|t} &=& b_{\tau+1|t} + r_\tau l_{\tau|t} + \expect_t R_\tau(w_\tau) l_{\tau|t}^2 + 
\expect_t[ c(w_\tau)D_{\tau+1|t} c(w_\tau)] + \nonumber \\
&&d_{\tau+1|t}^T(e_1 l_{\tau|t} + \expect_t c(w_\tau)) + 
l_{\tau|t}e_1^T D_{\tau+1|t} (e_1 l_{\tau|t} + 2 \expect_t c(w_\tau)) = \nonumber\\
&&b_{\tau+1|t} + \expect_t[ c(w_\tau)D_{\tau+1|t} c(w_\tau)] + d_{\tau+1|t}^T\expect_t c(w_\tau)
\label{eqnOptimalValue:const}
\EEAS
for $\tau = t, \ldots, T-1$.

\subsection{SHDP simplied solution (without value function)}
\label{appendix:shdp_solution_formulas}
Parts of the equations derived in Appendix \ref{appendix:shdp_solution_formulas_base}
are superfluous in case we are not interested in the
cost-to-go functions $v_{\tau|t}(x_t)$ for $\tau = t, \ldots, T-1$.
(In fact, we only want to compute 
the optimal action \eqref{eq:def_dynamic_solution_SHDP}.)
We disregard the constant term $b_{\tau|t}$,
and we only compute the three scalar elements that we need 
from $D_{\tau|t}$ and $d_{\tau|t}$. For any 
$t = 1, \ldots, T$
and $\tau = t, \ldots, T-1$ we define
\BEAS
e_1^t D_{\tau|t} e_1 &\equiv& \beta_{\tau|t} \\
e_1^t D_{\tau|t} e_2 = e_2^t D_{\tau|t} e_1 &\equiv&  \gamma_{\tau|t} \\
e_1^t d_{\tau|t} &\equiv& \delta_{\tau|t} \\
\EEAS
where $e_1 = (1,0)$ and $e_2 = (0, 1)$ are the unit vectors.
The final values are
\BEAS
\beta_{T|t} &=& \lambda \frac{\sigma_T^2}{C^2} + \frac{\alpha  s_T}{2C}\expect_t[{1}/{m_T}] \\
\gamma_{T|t} &=& - \lambda \frac{\sigma_T^2}C \expect_t[1/V] \\
\delta_{T|t} &=& \frac{ s_T}{2C} - {\alpha  s_T}\expect_t[{1}/{m_T}].
\EEAS
The policy 
\BEAS
K_{\tau|t} &=& - \frac{(\beta_{\tau+1|t}, \gamma_{\tau+1|t})}
{({\alpha  s_\tau}/{2C})\expect_t[{1}/{m_\tau}] + \beta_{\tau+1|t}} \\
l_{\tau|t} &=& - \frac{-{ s_\tau}/{(2C)} + \delta_{\tau+1|t} + 2\gamma_{\tau+1|t}\Expect_t m_\tau }
{2(({\alpha  s_\tau}/{2C})\expect_t[{1}/{m_\tau}] +  \beta_{\tau+1|t})}.
\EEAS
We restrict the Riccati equations 
to these three scalars.
They are independent from the rest of the recursion and we obtain
\BEAS
\beta_{\tau|t} &=& \lambda \frac{\sigma_\tau^2}{C^2} - \frac{\beta^2_{\tau+1|t}}
{({\alpha  s_\tau}/{2C})\expect_t[{1}/{m_\tau}]  + \beta_{\tau+1|t}}
+ \beta_{\tau+1|t}  =\\
&& \lambda \frac{\sigma_\tau^2}{C^2} + \frac{({\alpha  s_\tau}/{2C})\expect_t[{1}/{m_\tau}] \beta_{\tau+1|t}}
{({\alpha  s_\tau}/{2C})\expect_t[{1}/{m_\tau}]  + \beta_{\tau+1|t}}\\
\gamma_{\tau|t} &=& - \lambda \frac{\sigma_\tau^2}{C} \expect_t\left[1/{V}\right] -
\frac{\beta_{\tau+1|t} \gamma_{\tau+1|t}}{({\alpha  s_\tau}/{2C})\expect_t[{1}/{m_\tau}]  + \beta_{\tau+1|t}}
+ \gamma_{\tau+1|t} = \\
&&
 - \lambda \frac{\sigma_\tau^2}{C} \expect_t\left[1/{V}\right] + 
\frac{({\alpha  s_\tau}/{2C})\expect_t[{1}/{m_\tau}] \gamma_{\tau+1|t}}
{({\alpha  s_\tau}/{2C})\expect_t[{1}/{m_\tau}]  + \beta_{\tau+1|t}} \\
\delta_{\tau|t} &=& \delta_{\tau+1|t} + 2 \beta_{\tau+1|t}l_{\tau|t} + 
2\gamma_{\tau+1|t} \expect_t m_t. 
\EEAS

\subsubsection{Negligible spread}
\label{appendix:subsubsec:neglig_spread}
We study the case where 
$s_t = 0$ for all $t=  1, \ldots, T$, 
 equivalent to the limit $\lambda \to \infty$. 
From the equations above we get that for all $t=  1, \ldots, T$ and $\tau = t, \ldots, T$
\BEAS
\beta_{\tau|t} &=& \lambda \frac{\sigma_\tau^2}{C^2}  \\
\gamma_{\tau|t} &=& -\lambda \frac{\sigma_\tau^2}{C} \expect_t[1/V] \\
\delta_{\tau|t} &=& 0.
\EEAS
So for every $t =1, \dots, T$
\begin{multline*}
\label{eq:solution_shdp_lambda_infty}
\mu_{t|t}(x_t) = K_{t|t} x_t + l_t = \frac{-(\beta_{\tau+1|t}, \gamma_{\tau+1|t}) x_t - 
\expect_t m_t\gamma_{\tau+1|t}} {\beta_{t+1|t}}=\\
 C \expect_t \left[1/{V}\right] \left(\sum_{\tau = 1}^{t-1} m_\tau + \expect_t m_t\right) - \sum_{\tau = 1}^{t-1} u_\tau.
\end{multline*}
In other words, at every point in time we look at the difference between the fraction of order volume we have executed
and the fraction of daily volume the market has traded (using our most recent estimate of the total volume). We trade the 
expected fraction for next period $C \expect_t \left[1/{V}\right] \expect_t m_t$, plus this difference.

\section{Volume model}
\label{appendix:volume_model} 
We explain here the details of the volume model \eqref{eqDefVolumeModel},
which we use for the dynamic VWAP solution. In \S\ref{subsec:appendix:vol_estimation}
we describe the \emph{ad hoc} procedure we use to estimate the parameters of the model
on historical data. Then in \S\ref{subsec:appendix:volume_cross_val} we detail
the cross-validation of a particular feature of the model. Finally in \S\ref{subsec:appendix:expected_values_volume}
we derive formulas for the expected values of some functions of the volume, which
we need for the solution \eqref{eq:def_dynamic_solution_SHDP}.

\subsection{Estimation on historical data}
\label{subsec:appendix:vol_estimation}
We consider estimation of the volume model parameters $b^{(k)}$, 
$\mu$ and $\Sigma$ using data from days $i-W, \ldots, i-1$ 
(we are solving the problem at day $i$).
We append the superscript $(i,k)$ to any
quantity that refers to market day $i$ and stock $k$.

\paragraph{Estimation of $b^{k}$}
We first estimate the value of $b^{(k)}$ for each stock $k$, as:
\[
\hat{b}^{(k)} = \frac{
\sum_{j = i -W}^{i-1}
\sum_{t=1}^T \log m_t^{(j, k)}}{TW}
\]
We show in Table \ref{tableBvalues} the values of $\hat{b}^{(k)}$ obtained on
the first $W$ days of our dataset. 

\begin{table}[htb]
\begin{center}
\begin{tabular}{r||lr||l}
Stock & $\hat{b}^{(k)}$ & Stock & $\hat{b}^{(k)}$ \\
\hline
AA    & 4.338  & JPM   & 4.599 \\
AXP   & 3.910  & KO    & 4.312 \\
BA    & 3.845  & MCD   & 4.017 \\
BAC   & 5.309  & MMM   & 3.701 \\
CAT   & 4.118  & MRK   & 4.176 \\
CSCO  & 4.693  & MSFT  & 4.848 \\
CVX   & 3.986  & PFE   & 4.586 \\
DD    & 3.990  & PG    & 4.088 \\
DIS   & 4.055  & T     & 4.566 \\
GE    & 4.784  & TRV   & 3.546 \\
HD    & 4.139  & UNH   & 3.902 \\
HPQ   & 4.577  & UTX   & 3.782 \\
IBM   & 3.788  & VZ    & 4.225 \\
INTC  & 4.860  & WMT   & 3.992 \\
JNJ   & 4.244  & XOM   & 4.260 \\ 
\end{tabular}
\end{center}
\caption{
Empirical estimate $\hat{b}^{(k)}$ of the per-stock component
of the volume model, using data from the first $W = 20$ days.}
\label{tableBvalues}
\end{table}

\begin{figure}
\centerline{\includegraphics[width=20cm]{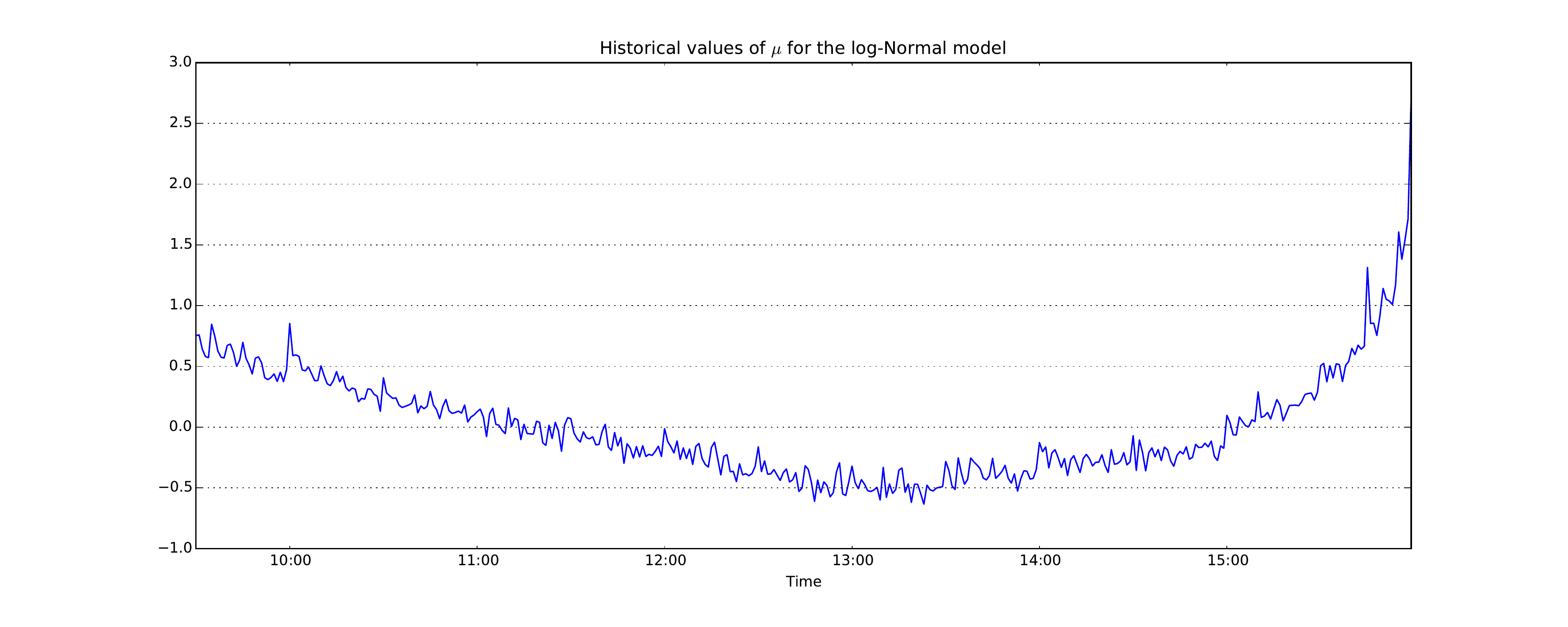}}
\caption{Empirical estimate $\hat \mu$ of the cross-time component
of the volume model, using data from the first $W = 20$ days.}
\label{fig:mu_unregularized}
\end{figure}

\paragraph{Estimation of $\mu$}
Since each observation $\log m^{(j,k)} - \ones b^{(j,k)}$ is distributed 
as a multivariate Gaussian we use this empirical mean as estimator of $\mu$:
\[
\hat{\mu}_t = \frac{
\sum_{j = i -W}^{i-1}
\sum_{k = 1}^K
\log m^{(j,k)}_t - \hat b^{(k)}}{WK}.
\]
We plot in Figure (\ref{fig:mu_unregularized}) the value of $\hat{\mu}$ obtained on
the first $W$ days of our dataset. 

\begin{figure}
\centerline{\includegraphics[width=14cm]{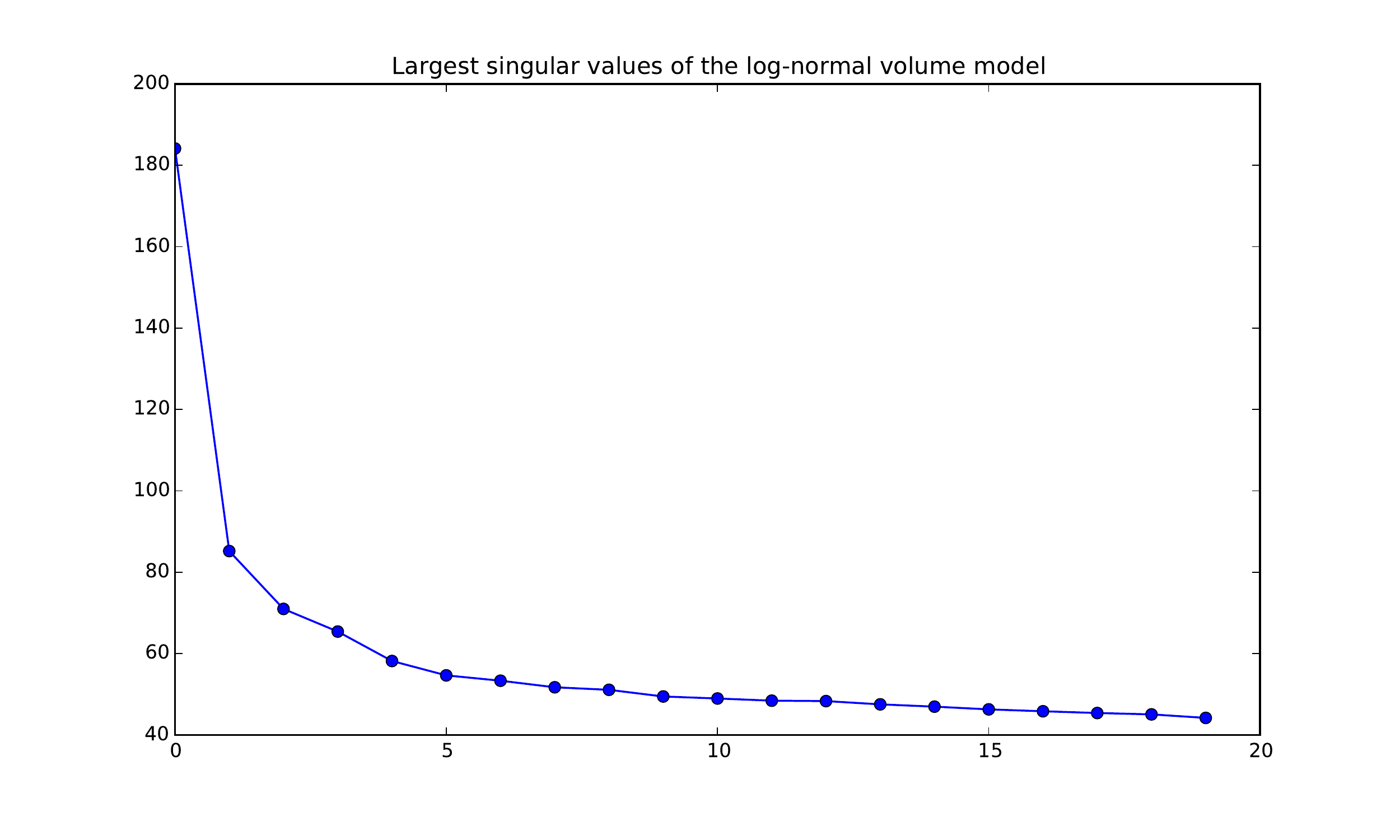}}
\caption{First 20 singular values of the matrix $X$ of observations
 $\log m^{(j,k)} - \ones \hat b^{(k)} - \hat{\mu}$.}
\label{fig:largest_SVs}
\end{figure}

\paragraph{Estimation of $\Sigma$}
We finally turn to the estimation of the covariance matrix $\Sigma \in \symm^T_{++}$,
using historical data. In general,
empirical estimation of covariance matrices is a complicated problem. Typically
one has not access to enough data to avoid overfitting (a covariance
matrix has $O(N^2)$ degrees of freedom, where $N$ is the dimension of a sample).
Many approximate approaches have been developed in the econometrics and statistics literature. 
We designed an \emph{ad hoc} procedure, inspired by works such as \cite{fan2011high}. We look
for a matrix of the form
\[
\Sigma = ff^T + S,
\]
where $f \in \reals^{T}$ and $S \in \symm^T_{++}$ is sparse. 
We first build the empirical covariance matrix.
Let $X \in \reals^{T\times {(WK)}}$ be the matrix whose columns are vectors of the form:
\[
	\log m^{(j,k)} - \ones \hat b^{(k)} - \hat{\mu}
\]
for each day $j = i-W, \ldots, i-1$ and stock $k = 1, \ldots, K$.
Then the empirical covariance matrix is
\[
\hat{\Sigma} = \frac{1}{WK-1} X X^T.
\]
We perform the singular value decomposition of $X$
\[
	X = U \cdot \diag({ s_1, s_2, \ldots, s_T}) \cdot V^T,
\]
where
$s\in \reals^T$, $s_1 \geq s_2 \geq ...\geq s_T \geq 0$, 
$U \in \reals^{T\times T}$, and $V \in \reals^{(WK)\times T}$
(because in practice we have $WK > T$, since $W=20$, $K=30$, and $T=390$).
We have
\[
\hat{\Sigma} = \frac{1}{WK-1} U \cdot \diag(s_1^2, s_2^2, \ldots, s^2_T) \cdot U^T.
\]
We show in \figref{fig:largest_SVs} the first singular values $s_1, s_2, \ldots, s_{20}$ computed on 
data from the first $W$ days. 
It is clear that the first singular value is much larger than all the others. We thus build
the rank 1 approximation of the empirical covariance matrix by keeping the first singular value
and first (left) singular vector 
\[
f = \frac{s_1 U_{:,1}}{\sqrt{WK-1}},
\] 
so that $ff^T$ is the best (in Frobenius norm) rank-1 approximation of $\hat \Sigma$.
We now need to provide an approximation for the sparse part $S$ of the covariance matrix.
We assume that $S$ is a \emph{banded} matrix of bandwidth $b > 0$,
which is non-zero only on the main diagonal and on $b-1$ diagonals above and below it
(in total it has $2b - 1$ non-zero diagonals).
The value of $b$ is chosen by cross-validation,
as explained in \S\ref{subsec:appendix:volume_cross_val}.
The assumption that $S$ is banded is inspired by
the intuition that elements of 
$\log m^{(j,k)} - \ones b^{(k)} - \mu$ are correlated (in time)
for short delays.
We find $S$ by simply copying the diagonal
elements of the empirical covariance matrix:
\[
\cdot S_{i,j} = \begin{cases}
(\hat{\Sigma} - ff^T)_{i,j} & \mbox{ if } |j -i| \leq b \\
0 &  \mbox{ otherwise. } \\
\end{cases}
\] 
We thus have built a matrix of the form 
$
{\Sigma} = ff^T + S.
$
Note that this procedure does not guarantee that ${\Sigma}$ is positive
definite. However in our empirical tests we always got 
positive definite ${\Sigma}$ for any $b = 1, 2 ,\ldots$.

\subsection{Cross validation}
\label{subsec:appendix:volume_cross_val}
\begin{figure}
\centerline{\includegraphics[width=11cm]{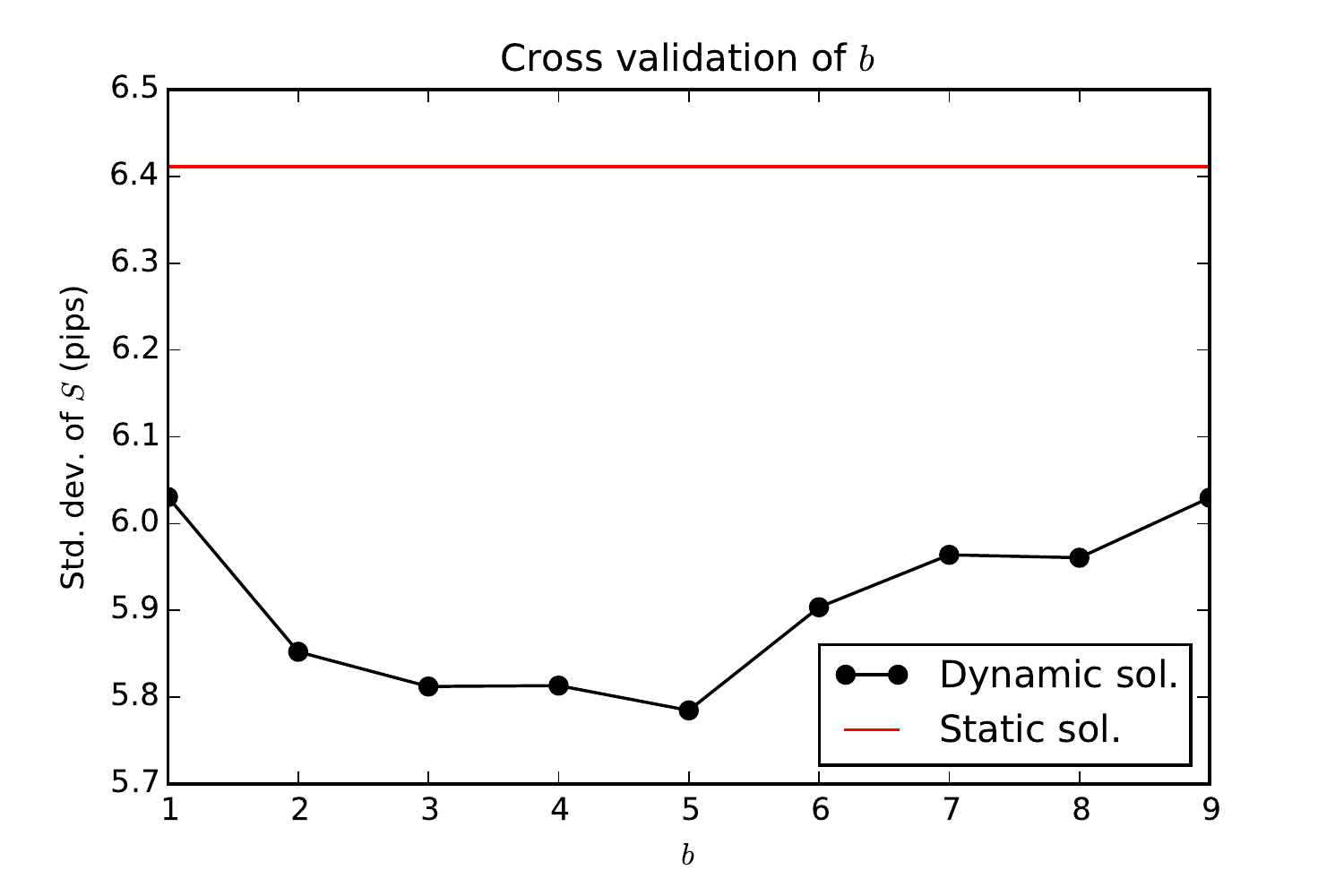}}
\caption{To cross validate the volume model
parameter $b$, we compute the empirical standard deviation of $S$ for the dynamic solution 
\eqref{eq:def_dynamic_solution_SHDP} 
  with $\lambda = \infty$, changing the value of $b$ in the volume model.
We also show the static solution 
\eqref{eq:solutiona_static_no_spread}, which does not use the volume model, for comparison.
From this result we choose $b = 3$ (to avoid overfitting).}
\label{fig:cross_validation_b}
\end{figure}

As explained in \S\ref{subsec:appendix:vol_estimation}, we need
to choose the value of the parameter $b \in \mathbb{N}$ (used
 for empirical estimation of the covariance matrix $\Sigma$).
 We choose it by cross-validation, reserving the first $W_{CV} = 10$ testing
 days of the dataset. We show
  in \figref{fig:rolling_testing_schematic} the way we partition the data 
  (so that the empirical testing is performed out-of-sample with respect to
  the cross-validation).
  We simulate trading according to the solution \eqref{eq:def_dynamic_solution_SHDP} 
  with $\lambda = \infty$
(\ie, the special case of Appendix \ref{appendix:subsubsec:neglig_spread}), 
for various values of $b$. We then compute the empirical variance of $S$, 
and choose the value of $b$ which minimizes it.
(We are mostly interested in optimizing the variance of $S$,
rather than the transaction costs.)
In \figref{fig:cross_validation_b} we show the result of this procedure (we show the
standard deviations instead of variances, for simplicity), along with the
result using the static solution \eqref{eq:solutiona_static_no_spread},
for comparison. Since the difference in performance
between $b=3$ and $b=5$ is small (and we want to avoid overfitting),
we choose $b = 3$. 

\subsection{Expected values of interest}
\label{subsec:appendix:expected_values_volume}
We consider the problem at any fixed time $t = 1, \ldots, T-1$,
for a given stock $k$ and day $i$.
(We have observed
market volumes $m_1, \ldots, m_{t-1}$.) 
We obtain the conditional distribution of the 
unobserved volumes $m_t, \ldots, m_T$ and derive expressions for
$\expect_t m_\tau$, $\expect_t \left[\frac{1}{m_\tau}\right]$,
and $\expect_t \left[\frac{1}{V}\right]$ for any $\tau = t, \ldots, T$. We need these for
the numerical solution \eqref{eq:def_dynamic_solution_SHDP},
as developed in Appendix \ref{appendix:subsubsec:neglig_spread}.

\paragraph{Conditional distribution}
We divide the covariance matrix in blocks:
\[
\Sigma = \left(
\begin{array}{cc}
\Sigma_{1:(t-1), 1:(t-1)} & \Sigma_{1:(t-1), t:T} \\
\Sigma_{t:T, 1:(t-1)} & \Sigma_{t:T, t:T}
\end{array}
\right).
\]
Then we get the marginal distribution
\[
m_{t:T} \sim 
\log \mathcal{N}(\nu{|{t}}, \Sigma{|{t}})
\]
by taking the \emph{Schur complement} (\eg, \cite{boyd2009convex}) of the covariance matrix 
\BEAS
\nu{|{t}} &\equiv& \mu_{{t}:T} + b^{(k)} + \Sigma_{1:(t-1), t:T}^T\Sigma_{1:(t-1), 1:(t-1)}^{-1}(\log m_{1:({t}-1)} - \mu_{1:({t}-1)} - b^{(k)}) \\
\Sigma{|{t}} &\equiv& \Sigma_{ t:T, t:T} - \Sigma_{1:(t-1),  t:T}^T\Sigma_{1:(t-1), 1:(t-1)}^{-1}\Sigma_{1:(t-1), t:T}.
\EEAS
 Note that $\nu{|1} = \mu + b^{(k)}$ and $\Sigma{|1} = \Sigma$, \ie, the unconditional distribution
of the market volumes.
We now develop the conditional expectation expressions.

\paragraph{Volumes}
The expected value of the remaining volumes $m_\tau$
\[
\Expect_t m_\tau = \exp\left((\nu{|t})_{\tau - t + 1} + \frac{(\Sigma|t)_{\tau - t + 1, \tau - t + 1}}{2}\right),
\quad \tau = t, \ldots, T.
\]
(Because the $(\tau - t + 1)$-th element of $\nu{|t}$ corresponds to the $\tau$-th volume.)

\paragraph{Inverse volumes}
The expected value of the inverse of the remaining volumes $m_\tau$
\[
\Expect_t \left[\frac{1}{m_\tau}\right] = \exp\left({-(\nu{|t})_{\tau - t + 1} + \frac{(\Sigma|t)_{\tau - t + 1, \tau - t + 1}}{2}}\right),
\quad  \tau = t, \ldots, T.
\]

\paragraph{Total volume}
We have, since we already observed $m_1, \ldots, m_{t-1}$
\[
\Expect_t V = \sum_{\tau=1}^{t-1} m_\tau + \sum_{\tau =t }^T \Expect_t m_\tau.
\]
We also express its variance, which we need later
\begin{multline*}
\var_t(V) = \var_t \sum_{\tau =t }^T m_\tau = \sum_{\tau =t }^T \sum_{\tau' =t }^T \mathbf{cov}(m_\tau, m_{\tau'}) = \\
\sum_{\tau =t }^T \sum_{\tau' =t }^T
\Expect_t m_\tau \Expect_t m_{\tau'} \left(\exp((\Sigma|t)_{\tau - t + 1, \tau' - t + 1}) -1 \right).
\end{multline*}

\paragraph{Inverse total volume}
We use the following approximation, derived from the Taylor expansion formula.
Consider a random variable $z$ and a smooth function $\phi(\cdot)$, then
\[
\expect \phi(z) \simeq \phi (\expect z) + \frac{\phi''(\expect z)}{2}\var z.
\]
So the inverse total volume
\[
\expect_t\left[\frac{1}{V}\right] \simeq \frac{1}{\expect_t{V}} + \frac{\var_t(V)}{\expect_t[{V}]^3}.
\]

\end{appendices}

\bibliography{../biblioVWAParticle}{}
\bibliographystyle{plain}
\end{document}

%% file: defs.tex
\newcommand{\BEAS}{\begin{eqnarray*}}
\newcommand{\EEAS}{\end{eqnarray*}}
\newcommand{\BEA}{\begin{eqnarray}}
\newcommand{\EEA}{\end{eqnarray}}
\newcommand{\BEQ}{\begin{equation}}
\newcommand{\EEQ}{\end{equation}}
\newcommand{\BIT}{\begin{itemize}}
\newcommand{\EIT}{\end{itemize}}
\newcommand{\BNUM}{\begin{enumerate}}
\newcommand{\ENUM}{\end{enumerate}}

\newcommand{\BA}{\begin{array}}
\newcommand{\EA}{\end{array}}

\newcommand{\eg}{{\it e.g.}}
\newcommand{\ie}{{\it i.e.}}

\newcommand{\ones}{\mathbf 1}

\newcommand{\reals}{{\mbox{\bf R}}}
\newcommand{\integers}{{\mbox{\bf Z}}}

\newcommand{\symm}{{\mbox{\bf S}}}  

\newcommand{\diag}{\mathop{\bf diag}}

\newcommand{\Expect}{\mathop{\bf E{}}}
\newcommand{\expect}{\mathop{\bf E{}}}

\newcommand{\argmin}{\mathop{\rm argmin}}

\newcommand{\var}{\mathop{\bf var}}

{\begin{quote}}{\end{quote}}

\makeatletter
\long\def\@makecaption#1#2{
   \vskip 9pt 
   \begin{small}
   \setbox\@tempboxa\hbox{{\bf #1:} #2}
   \ifdim \wd\@tempboxa > 5.5in
        \begin{center}
        \begin{minipage}[t]{5.5in}
        \addtolength{\baselineskip}{-0.95pt}
        {\bf #1:} #2 \par
        \addtolength{\baselineskip}{0.95pt}
        \end{minipage}
        \end{center}
   \else 
	\hbox to\hsize{\hfil\box\@tempboxa\hfil}  
   \fi
   \end{small}\par
}
\makeatother

\newcounter{oursection}

\newcounter{lecture}